\documentclass[structabstract]{aa}  

\usepackage{graphicx}
\usepackage{txfonts}

\begin{document}

   \title{Polytropic neutron star -- black hole merger simulations
     with a Paczy\'nski-Wiita potential}

   \titlerunning{Polytropic neutron star -- black hole merger simulations}

   \author{M.~Ruffert\inst{2,1}
          \and
          H.-Th.~Janka\inst{1}
          }

   \institute{ Max-Planck-Institut f\"ur Astrophysik, 
               Karl-Schwarzschild-Str.~1, D-85748 Garching bei
               M\"unchen, Germany\\
               \email{thj@mpa-garching.mpg.de}
         \and  The School of Mathematics and the Maxwell Institute, 
               University of Edinburgh, King's Buildings,
               Edinburgh EH9 3JZ, U.K.\\
               \email{m.ruffert@ed.ac.uk}
               }


\abstract
{Mergers of neutron stars (NS) and black holes (BH) are among the
  strongest sources of gravitational waves and are potential central
  engines for short gamma-ray bursts.}  
{We aim to compare the general relativistic (GR) results of other
  groups with Newtonian calculations of models with equivalent
  parameters. We vary the mass ratio of the NS to BH and the
  compactness of the NS. The mass of the NS is $1.4\,M_\odot$. 
  We compare the dynamics in the parameter-space regions where the NS
  is expected to reach the innermost stable circular orbit (ISCO)
  before being tidally disrupted (mass shedding, MS), and vice versa.}  
{The hydrodynamics is evolved by a Newtonian PPM scheme with four
  levels of nested grids. We
  use a polytropic EoS ($\Gamma\!=\!2$), as adopted in the GR
  simulations. However, instead of full GR we use a Newtonian
  potential supplemented by a Paczy\'nski-Wiita-Artemova potential for
  the BH, both disregarding and including rotation of the BH.} 
{If the NS is compact (${\cal C}\!=\!0.18$), it is accreted by the BH more
  quickly, and only a small amount of mass remains outside the BH. If
  the mass ratio 
  is small ($Q\!=\!2$ or 3) or the NS is less compact (${\cal C}\!=\!0.16$ or
  less), the NS is tidally torn apart before being accreted. Although
  most of the mass is absorbed by the BH, some 0.1~$M_\odot$ remain
  in a tidal arm. For small mass ratios ($Q\!=\!2$ and 3), the tidal
  arm can wrap around the BH to form a thick disk. When including the
  effects of either BH spin-up or spin-down by the accreted matter,
  more mass remains in the surroundings (0.2--0.3~$M_\odot$).} 
{Although details and quantitative results differ, the general trends
  of our Newtonian calculations are similar to the GR
  calculations. A clear delimiting line separating the ISCO from the
  MS cases is not found. Inclusion of BH rotation as well as
  sufficient numerical resolution are extremely important.}


\maketitle


\section{Introduction}

Merging binaries of neutron stars (NS) and black holes (BH) are of
interest both as possible central engines for short gamma-ray bursts
and as promising gravitational wave sources detectable by the
interferometer observatories. 
As computational power and numerical schemes have progressed so has
the detail of simulations of these mergers.  
Early simulations by our group (Janka et al.~1999) were developed to
investigate gamma-ray burst scenarios. 
We included detailed microphysics (equation of state, neutrino
source terms) but accounted for general
relativistic effects either only partially (gravitational wave
emission and backreaction) or phenomenologically (Paczy\'nski-Wiita
potential for the BH).  
Other groups interested in gravitational wave aspects have published
results of NS--BH mergers with general relativistic (GR) physics 
included to different levels of approximation
(Etienne et al.~2009; Shibata et al.~2009;
Rantsiou et al.~2008; Taniguchi et al.~2008; Duez et al.~2008), but
with some variant of the polytropic equation-of-state (EoS) for the
neutron star matter. 
Rantsiou et al.~(2008) did not include dynamic GR in that they
used a fixed background. Ferrari et al.~(2009) investigated
quasi-equilibrium sequences of ellipsoidal NSs also including
microphysical EoSs in the tidal approximation. A good source of
references are the reviews of Faber (2009) and Duez (2009).

\begin{table*}
\caption{Key initial model parameters and some results.} 
\label{tab:models}      
\centering          
\begin{tabular}{c c c c c c r l c l c c } 
\hline\hline       
Model & $Q$ & $\cal{C}$ & $M_{\mathrm{BH}}$ & 
$R_{\mathrm{NS}}$ & $d_{\mathrm{i}}$ 
 & $L$ & $\Delta x$ & $M_{\mathrm{u}}$ 
 & $M_{\mathrm{b}}$ & $M_{\mathrm{g}}$  & $|$h/l$|$ \\ 
      &     &           & $M_\odot$ & km & km 
 & km & km & $M_\odot$  & $M_\odot$ & $M_\odot$  &  \\ 
\hline
 M5.145 & 5 & 0.145 & 7.0 & 14.3 & 92 & 1550 & 0.77 & 0.07 & 0.08 
  &  0.15     & N  \\ 
 M4.145 & 4 & 0.145 & 5.6 & 14.3 & 80 & 1400 & 0.68 & 0.08 & 0.10 
  & 0.18      & Y  \\
 M3.145 & 3 & 0.145 & 4.2 & 14.3 & 68 & 1100 & 0.54 & 0.06 & 0.09 
  & 0.15      & Y \\
 M2.145 & 2 & 0.145 & 2.8 & 14.3 & 56 &  880 & 0.43 & $<10^{-2}$ & 0.07 
  & 0.08      & Y \\  
 M5.160 & 5 & 0.160 & 7.0 & 12.9 & 92 & 1550 & 0.77 & 0.01 & 0.05 
  & 0.06      &  -  \\
 M4.160 & 4 & 0.160 & 5.6 & 12.9 & 76 & 1250 & 0.61 & 0.03 & 0.06 
  & 0.09      &  N  \\
 M3.160 & 3 & 0.160 & 4.2 & 12.9 & 66 & 1100 & 0.54 & 0.03 & 0.06 
  & 0.09      &  N  \\
 M2.160 & 2 & 0.160 & 2.8 & 12.9 & 52 &  820 & 0.40 & $<10^{-2}$ & 0.04 
  & 0.05       & Y  \\  
 M5.180 & 5 & 0.180 & 7.0 & 11.5 & 88 & 1500 & 0.73 & $<10^{-3}$ & $<10^{-3}$
  & $<10^{-3}$ & N \\
 M4.180 & 4 & 0.180 & 5.6 & 11.5 & 76 & 1250 & 0.61 & $<10^{-3}$ & $<10^{-3}$
  & $<10^{-3}$ & N \\
 M3.180 & 3 & 0.180 & 4.2 & 11.5 & 64 & 1050 & 0.51 & $<10^{-2}$ & 0.01 
  & 0.01       & N  \\
 M2.180 & 2 & 0.180 & 2.8 & 11.5 & 50 &  780 & 0.38 & $<10^{-3}$ & 0.004
  & 0.004      & N  \\  
\hline                  
\end{tabular}
\begin{list}{}{}
\item[] Initial parameters: mass ratio $Q$, compactness $\cal{C}$,
  black hole mass 
  $M_{\mathrm{BH}}$, neutron star radius $R_{\mathrm{NS}}$, initial orbital
  distance $d_{\mathrm{i}}$, size of largest 
  grid $L$, size of finest zone $\Delta x = L/2048$. 
  The mass of the NS is $1.4\,M_\odot$ in all cases.
  Values at the end of the simulation, only for high-resolution models: 
  unbound+ejected gas mass $M_{\mathrm{u}}$, 
  bound neutron star mass around BH $M_{\mathrm{b}}$, 
  neutron star mass not instantly accreted by BH $M_{\mathrm{g}}$, 
   $|$h/l$|$ states whether the difference between values for
  $M_{\mathrm{g}}$ in low-resolution and high-resolution simulations
  is less than $20\%$ (Yes/No; `-' no low-res model computed). The
  spin-up of the BH by accretion of matter is not taken into account. 
\end{list} 
\end{table*}

We compare the results of a partially post-Newtonian calculation of
NS--BH mergers with published GR simulations, to address in particular
the following aspect: Miller (2005) pointed
out that during the NS--BH binary orbital inspiral, the innermost
stable circular orbit (ISCO) can be reached before the NS becomes
tidally disrupted; this was confirmed by Taniguchi et al.~(2008) and
Ferrari et al.~(2009).  
In this case, it is to be expected that most of the NS matter is
swallowed by the BH before the material can spread out and form a
massive torus. 
Since the ISCO is a purely general relativistic effect, its
consequences will not be seen in a purely Newtonian simulation. 
However, the Paczy\'nski-Wiita potential (Paczy\'nski \& Wiita 1980,
Abramowicz~2009) phenomenologically mimics the ISCO in a Newtonian
setting. 
So the question arises as to whether and to what extent one can
reproduce the results of the general relativistic simulations with
crude Newtonian approximations.  

Which of the two cases of NS reaching ISCO first or NS becoming tidally
disrupted first happens during the evolution will have a direct 
impact upon the amount of mass remaining in the torus around the BH.
This mass is supposed to provide the energy for the jet of the
short-duration gamma-ray bursts, so a lower mass should produce a
dimmer burst. 
Intrinsically, the tidal disruption --- called `mass shedding' (MS)
in previous papers --- should yield a more massive torus than the ISCO case. 
As shown in our previous investigations (Ruffert et al.~2001 and
references therein, Setiawan et al.~2006), fairly massive tori of  
at least several hundredths of a solar mass seem to be needed for the
scenario of a hot neutrino-emitting disk to function as a central engine
for gamma-ray bursts (Lee et al.~2005, Oechslin \& Janka 2006).
Thus the question of how the NS becomes tidally disrupted or
swallowed by the BH, and how much mass remains inside the torus is an
important one to explain gamma-ray bursts that needs to be
investigated in detail. The results of previously published GR
simulations still disagree with each other in this point.

Following the introduction above, this paper will describe the methods
in the next section and the results of the simulations in
Sect.~\ref{sect:result}. These results are then discussed
(Sect.~\ref{sect:discuss}) and the paper concludes with a summary
(Sect.~\ref{sect:conclude}).


\section{Methods\label{sect:method}}

\begin{figure}
   \centering
   \includegraphics[bb= 50 100 500 450,width=9.cm]{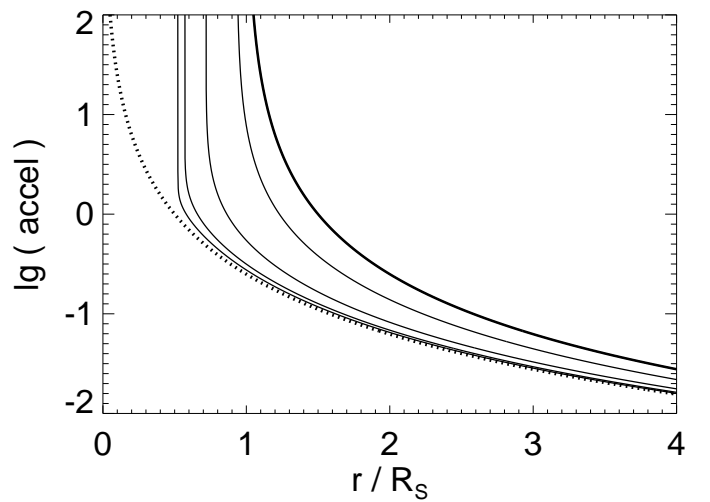}
   \caption{Gravitational acceleration as a function of distance
     $r$, as given by the prescription of Eq.~(\ref{eq:ArtePaWi}), for 
    various spin parameters $a=0.999$, 0.99, 0.9, 0.5, 0; solid lines
    from left to right. The dotted line shows a Newtonian potential 
    acceleration $\propto\!r^{-2}$. The bold solid line shows the 
    Paczy\'nski-Wiita case $\propto\!(r-R_{\mathrm{S}})^{-2}$,
    i.e.,~$a\!=\!0$. $R_{\mathrm{S}}$ is the Schwarzschild radius. }
    \label{fig:ArtePot}
\end{figure}

\subsection{Theory and numerical formulation}

The three-dimensional hydrodynamic simulations were performed with a
basically Newtonian code based on the piecewise parabolic method (PPM)
of Colella \& Woodward (1984) with four levels of nested grids 
(Ruffert 1992). 
The equidistant Cartesian grids all have the same number of zones,
depending on the model, either 128 or 256 zones per dimension. 
Initially the finest grid covers both components of the system
(neutron star and black hole).  
We use an ideal gas equation-of-state (EoS, $P\!=\!(\Gamma-1)\epsilon$
with adiabatic index $\Gamma\!=\!2$, pressure $P$, and internal energy
density $\epsilon$) to close the energy
equation evolved in the hydrodynamics code.
The self-gravity of the gaseous mass distribution on the grids are
calculated using fast-Fourier transforms (FFT). 
Details of the
implementation of the nested grids can be found in Ruffert~(1992).
A low density medium, of roughly $10^{-8}$ the central density of the
NS, fills the volume outside the NS and the BH. This ``atmosphere'' is 
necessary for numerical reasons (e.g., use of the reciprocal density)
and is common to grid-based codes, both for Newtonian and GR
simulations. The density of all zones is kept above this 
minimal value, which is referred to as ``vacuum''.

The black hole (BH) of mass $M_{\mathrm{BH}}$ is treated as a
gravitational point mass surrounded by a vacuum sphere discretised 
on the grids. 
The existence of the innermost stable circular orbit (ISCO) is
mimicked by including a Paczy\'nski-Wiita-type potential 
(Paczy\'nski \& Wiita 1980, Artemova et al.~1996). 
The Artemova et al.~prescription for the acceleration is
\begin{equation}
\frac{\mathrm{d}\Phi_{\mathrm{BH}}}{\mathrm{d}r} = 
    -\frac{GM_{\mathrm{BH}}}{r^{2-\beta}(r-r_{\mathrm{H}})^\beta} \,, 
\label{eq:ArtePaWi}
\end{equation}
where $r_{\mathrm{H}}$ is the black hole event horizon, and $\beta$ is
a constant for a given value of the BH spin parameter $a$. 
It is defined by  
$\beta(a)=(r_{\mathrm{in}}(a)-r_{\mathrm{H}}(a))/ r_{\mathrm{H}}(a)$, 
where $r_{\mathrm{in}}(a)$ is the radius of the ISCO. 
This prescription reduces to the usual Paczy\'nski-Wiita 
potential when the BH spin parameter $a$ is zero. We then have that 
$\beta=2$, and $r_{\mathrm{H}}(a=0)=2GM_{\mathrm{BH}}/c^2$. 
We note a mathematical discontinuity: for $\beta > 0$, $r_{\mathrm{H}}$
remains a lower limit for $r>r_{\mathrm{H}}$, whereas for $\beta=0$
one recovers the Newtonian case for all $r>0$ 
(see Fig.~\ref{fig:ArtePot}).
The graphs of $r_{\mathrm{H}}(a)$, $\beta(a)$, and $r_{\mathrm{in}}(a)$
can be found in Setiawan et al.~(2006), their Fig.~1.

This potential is implemented numerically as follows. 
First the total Newtonian potential is calculated for all gas on the
grids, including the BH mass. From this potential, the accelerations
are derived by discrete differencing, as is common procedure. 
We then add the acceleration given by the difference
between the formalism of Eq.~(\ref{eq:ArtePaWi}) and a purely
Newtonian potential. 
By construction, this difference only acts on
the gaseous NS matter on the grid. To respect Newton's Third Law
(``action=reaction''), a tab of the sum of these accelerations has to be
kept and its inverse enforced on the BH. 

The radius of the vacuum sphere mimicking the BH is chosen to be the
arithmetic mean of the event horizon and the ISCO. 
The mass, momentum, and angular momentum of material flowing across
this inner boundary are added to the BH values and the zones affected
are reset to near-vacuum values. 
The position of the BH is adjusted accordingly --- the centre-of-mass
of the gas leaving the grid and of the BH do not necessarily
coincide --- and moved by a leapfrog procedure (alternate position and
velocity updates). 
In all models listed in Table~\ref{tab:models}, the angular momentum of
the accreted material does not change the spin parameter $a$ in the
equations that determine the BH potential, Eq.~(\ref{eq:ArtePaWi});
i.e.,~the BH potential only deepens because of the increase in mass
of the BH. 
We evolved an additional set of four models to investigate the effects
of BH spin. These models are listed in Table~\ref{tab:Rot} and 
discussed in separate paragraphs.  

The general relativistic effect that is modelled
phenomenologically by the procedures described above is the presence
of the ISCO and its change in radius. 
Frame-dragging and many other effects, e.g.,~relativistic kinematics,
redshift, time dilation, are omitted. 
However, we do include the local effects of gravitational-wave
emission --- the volume integral of which reproduces the quadrupole
approximation --- and the corresponding back-reaction on the hydrodynamic
flow (for details see Ruffert et al.~1996).

\begin{figure}
   \centering
   \includegraphics[bb= 50 90 520 460,width=9.cm]{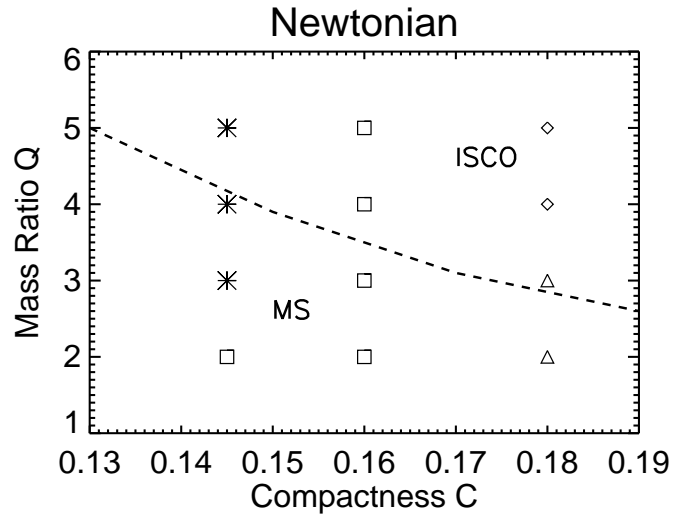}
   \caption{Parameter space of models presented in this paper
     (only non-spinning BH), with
     compactness ${\cal{C}}=(GM_{\mathrm{NS}})/(c^2R_{\mathrm{NS}})$ 
     and mass ratio $Q=M_{\mathrm{BH}}/M_{\mathrm{NS}}$. 
     Each symbol represents one model: 
     `diamond' $M_{\mathrm{g}}\!\!<\!\!10^{-3}M_\odot$,
     `triangle'  $10^{-3}M_\odot\!\!<\!\!M_{\mathrm{g}}\!\!<\!\!0.02M_\odot$,
     `square'  $0.02 M_\odot\!\!<\!\!M_{\mathrm{g}}\!\!<\!\!0.1M_\odot$,
     `star'  $0.1M_\odot\!<\!M_{\mathrm{g}}$, for the mass
     $M_{\mathrm{g}}$ not instantly accreted by the BH. Note that the
     spin-up of the BH by the accretion of matter is not taken into
     account. 
     The line separating the innermost stable circular orbit (ISCO) and
     the mass shedding (MS) regions is taken from Taniguchi et al.~(2008).}
   \label{fig:ISCO.MS}
\end{figure}

\begin{figure}
   \centering
   \includegraphics[bb= 50 90 520 460,width=9.cm]{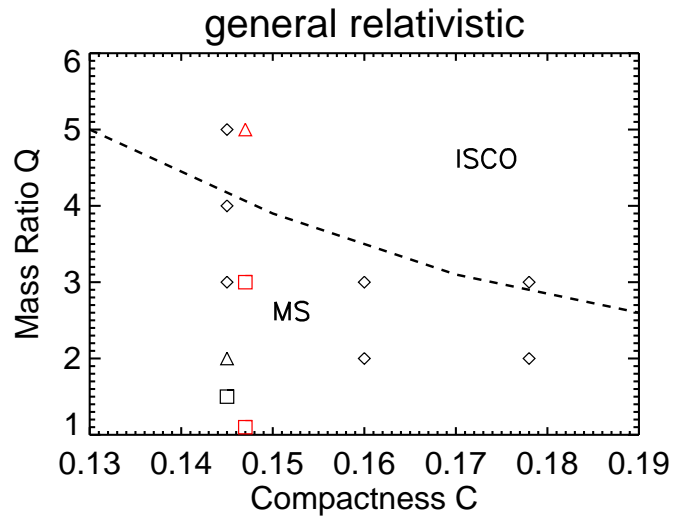}
   \caption{Results of Shibata et al.~(2009) and 
     Etienne et al.~(2009) (only non-spinning BH) models, with
     compactness ${\cal{C}}$ and mass ratio $Q$. Each symbol
     represents one model; we use the same symbols as used in 
     Fig.~\ref{fig:ISCO.MS}. The symbols for the Etienne et al.~(2009)
     models are slightly shifted to ${{\cal C}=0.147}$ (and coloured),
     for clarity. }  
   \label{fig:ISCO.MS.Shi}
\end{figure}

\subsection{Initial conditions\label{sect:initcond}}

The initial distance of NS to BH varies from model to model, to ensure
roughly 2--3 orbits before coalescence (see Table~\ref{tab:models}). 
For the finest grid to cover both components of the binary, the
geometric extent of the grids has to be adapted to ensure the
highest possible numerical resolution. 
Both the extent of the largest grid as well as the size of
the finest zone are listed in Table~\ref{tab:models}.

Parameters for the NS model are given as follows. The mass is kept
constant for all polytropic models, $M_{\mathrm{NS}}\!=\!1.4 M_\odot$.  
The radius, $R_{\mathrm{NS}}$, is then fixed by the chosen compactness, 
${\cal{C}}\!=\!(G M_{\mathrm{NS}})/(c^2 R_{\mathrm{NS}})$.  
For a $\Gamma\!=\!2$ polytrope, the relation between pressure $P$ and
density $\rho$ is $P=\kappa \rho^2$, 
where $\kappa=\frac{2}{\pi}G R_{\mathrm{NS}}^2$ is fixed by the NS radius.
The density distribution is $\rho(\xi) = \rho_c \frac{\sin(\xi)}{\xi}$,
with $M_{\mathrm{NS}} = \frac{4}{\pi} R_{\mathrm{NS}}^3 \rho_c$, 
and $\xi = \pi r / R_{\mathrm{NS}}$. 

The main variable of the BH, its mass, is defined in terms of its
mass ratio with respect to the NS,
$Q\!=\!M_{\mathrm{BH}}/M_{\mathrm{NS}}$. 
Both the NS and BH are initially given Keplerian velocities plus a
small radial velocity component from the quadrupole approximation of
gravitational wave emission. 

In our Newtonian framework, there is no ambiguity about which
masses and radii to use in the equations above. 
This differs from relativistic calculations, where a choice of
masses (e.g., ADM, rest), and distances (e.g., isotropic,
circumferential) exists. 
We return to this point when trying to compare our
results with other people's simulations, in Sect.~\ref{sect:discuss}.

When the NS is evolved in the simulation, the initially spherical NS
adapts to the BH potential, orbital motion, and grid resolution on its
(NS) dynamical timescale, i.e., within much less than one orbital
revolution.  
This initial phase is almost complete by the time the NS crosses the
ISCO or becomes tidally disrupted.  

The actual values of the mass ratio $Q$ and compactness $\cal{C}$ can
be found in Table~\ref{tab:models}. 
They are also shown in Fig.~\ref{fig:ISCO.MS} and were chosen to match
closely the models presented in Shibata et al.~(2009) (see 
Fig.~\ref{fig:ISCO.MS.Shi}).
Figures~\ref{fig:ISCO.MS} and~\ref{fig:ISCO.MS.Shi} include the line
separating the regions within which the NS first reaches the innermost
stable circular orbit (ISCO) and the mass shedding limit (MS),
respectively. 
It is taken from Taniguchi et al.~(2008) and is based on
quasi-equilibrium configurations of the binary system.


\begin{figure*}
   \centering
   \includegraphics[bb=30 100 565 765,clip,width=18.1cm]{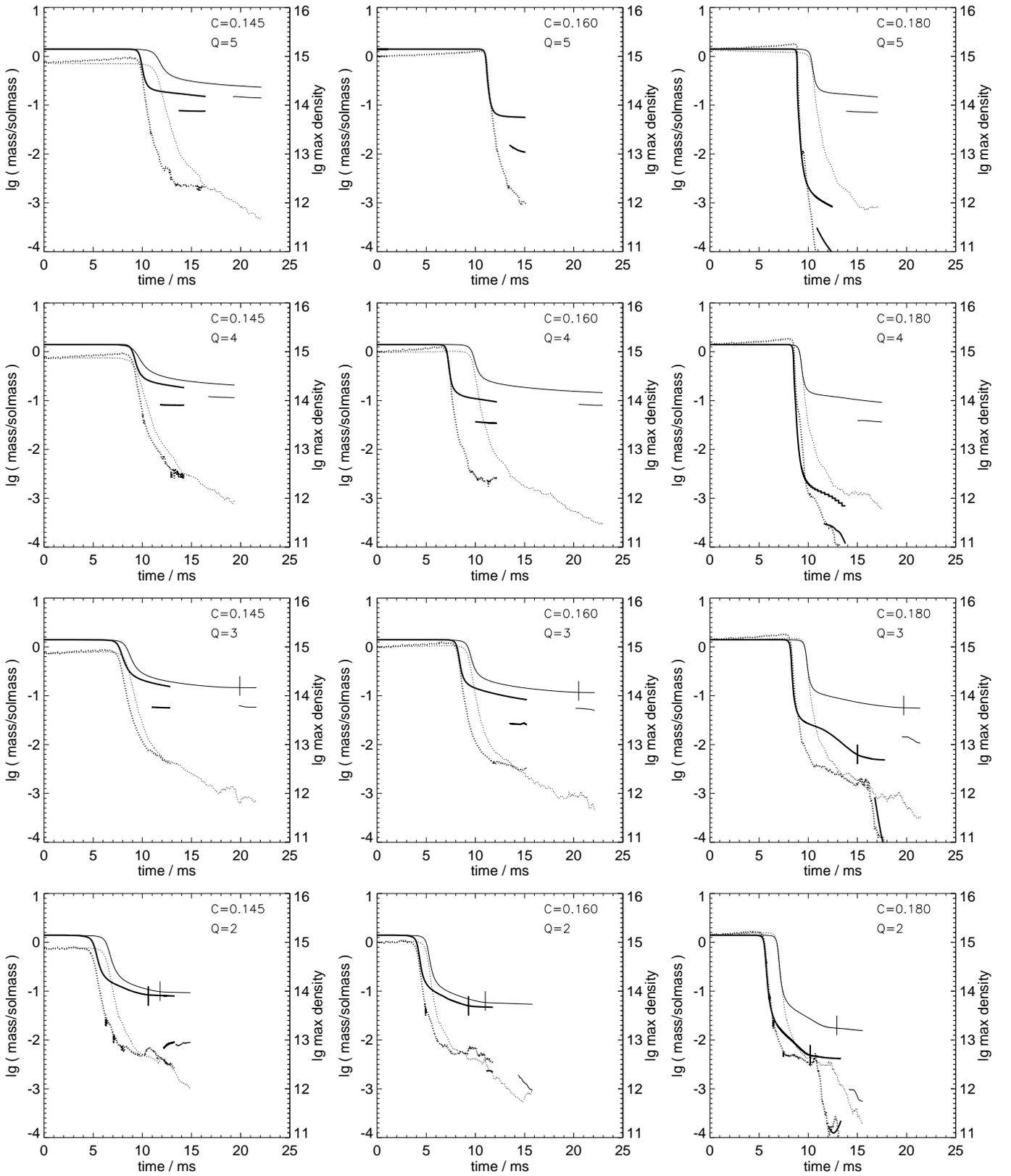}
   \caption{Temporal evolution of some quantities for various models;
     the parameter values for $Q$ and ${\cal C}$ are given within each
     panel. Solid lines show the gas mass. Dotted lines show the maximum
     density. Thin lines show low-resolution simulations ($128^3$);
     bold lines show high-resolution simulations ($256^3$).
     A short vertical line indicates the formation of an accretion
     disk. Short pieces of nearly horizontal lines show the amount of
     unbound mass.}  
   \label{fig:mass}
\end{figure*}

\begin{figure*}
   \centering
   \includegraphics[bb=21 95 539 755,clip,width=18.1cm]{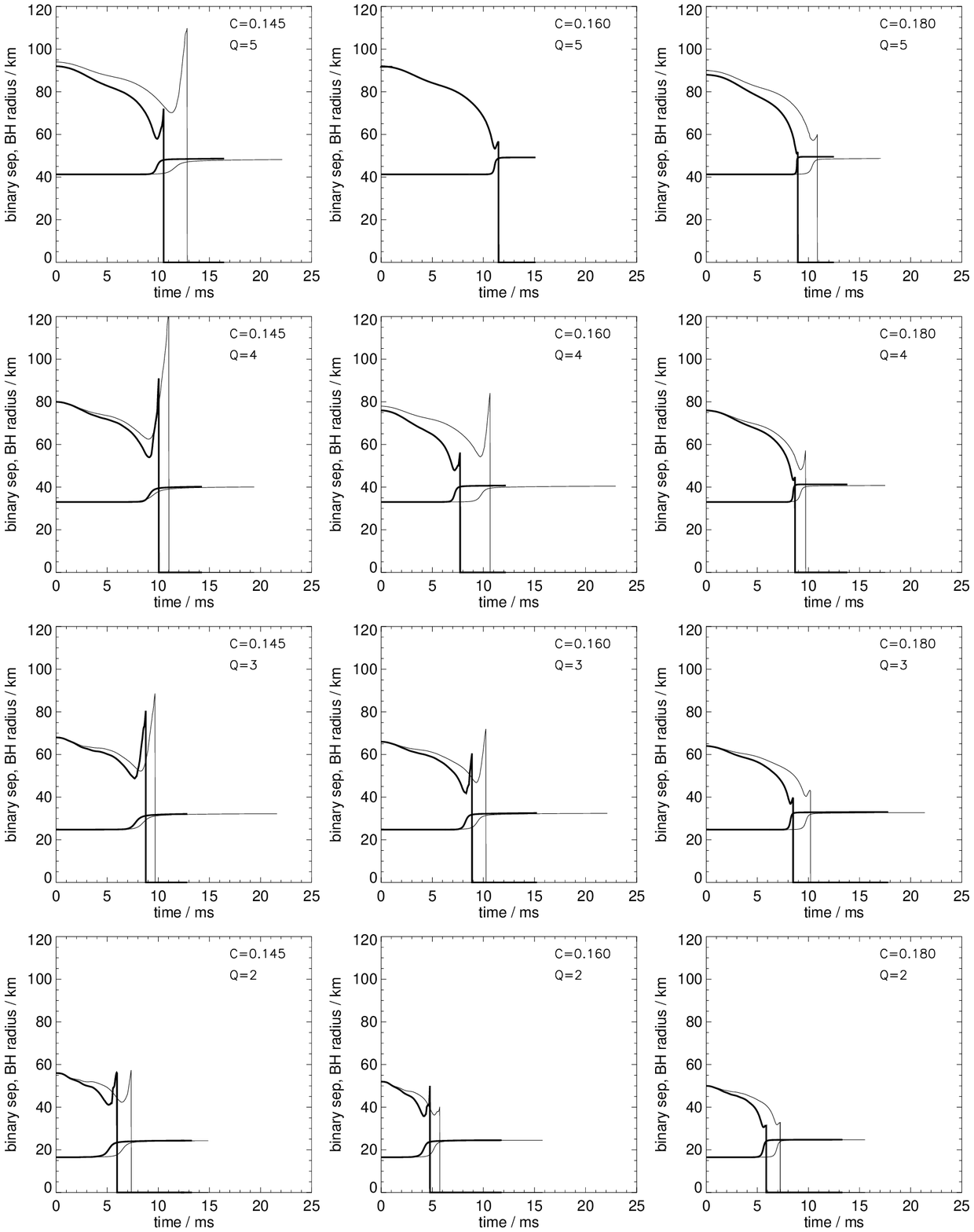}
   \caption{Temporal evolution of some quantities for various models;
     the parameter values for $Q$ and ${\cal C}$ are given within each
     panel. The upper pair of lines show binary separation; the lower
     pair of lines show the radius of the
     BH (details see text). Thin lines show low-resolution simulations
     ($128^3$); bold lines show high-resolution simulations ($256^3$).  
     The vertical line indicates when the mass within a sphere around
     the density maximum falls below $0.03\;M_\odot$.} 
   \label{fig:sep}
\end{figure*}


\section{Results\label{sect:result}}

\subsection{Resolution and convergence}

\begin{figure}
   \centering
   \includegraphics[bb= 50 90 520 440,width=9.cm]{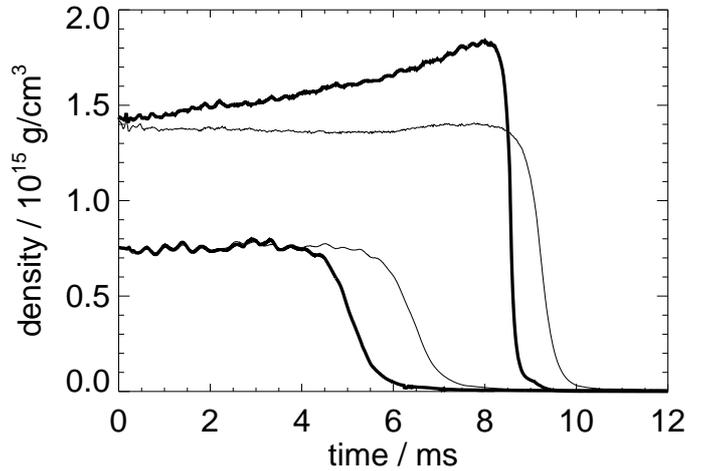}
   \caption{Maximum density as a function of time for two selected
     models. The upper pair of lines shows model M4.180; the lower
     pair of lines shows model M2.145. 
     Thin lines show lower-resolution simulations ($128^3$); bold
     lines show higher-resolution simulations ($256^3$).  } 
   \label{fig:densi}
\end{figure}

Figure~\ref{fig:mass} shows results for both high-resolution and
low-resolution models. 
Simulations with lower mass ratios and less compact NSs have
intrinsically better resolution (see Table~\ref{tab:models}). 
The dynamics for these models and the results, e.g., gas mass on the
grid after merger, agree reasonably well for the high- and
low-resolution calculations. 
Thus $128^3$ runs, which are computationally feasible in a few days,
yield converged results. 
However, the models with very compact NSs -- which have comparatively
fewer zones that cover the NS volume and in particular 
the steep surface gradients -- or models with a
high mass ratio -- i.e., a large BH, which in turn implies large zones
-- require the use of at least $256^3$ zones to be adequately resolved. 
This can also be seen in Fig.~\ref{fig:densi}, for example: 
for model M2.145, both the low- and high-resolution simulations can
hold the NS intact on the grid and thus the maximum density during the
first few orbits are equal. 
On the other hand, the maximum density of the NS in model M4.180 is
distinctly lower in the low-resolution simulation than in the
high-resolution calculation, which indicates that the NS develops a
more compact core during the initial orbital motion. For very low
resolution models, the NS core dissolves secularly.
For this model the results of the low-resolution simulation should be
interpreted with caution. 
We have no reason to believe that the high-resolution simulations are
compromised in this way, but are unable to perform even
higher-resolution calculations to check this point.

One notices that even for the simulations where the NS--BH
distance is initially identical, the high resolution models 
reach the point of merger systematically earlier than the low
resolution cases.  
This clearly numerical effect seems to decelerate the radial inward
motion of the lower resolution runs, even in the cases resolved best
($Q$=2) where the central densities for both resolutions match (model
M2.145 in Fig.~\ref{fig:densi}). A detailed analysis of the possible
cause(s) of this more rapid pre-merging evolution of more highly
resolved models did not provide an unambiguous conclusion. We could
identify differences in neither the gravitational-wave energy loss,
nor the quality of energy conservation, nor the precision of the
gravitational potential energy, as clear reason(s) for the observed
behaviour. It appears most likely that a difference in 
angular momentum conservation plays a role. This
quantity, however, is very difficult to calculate precisely with its
different contributions from gas, orbit, BH spin, and gravitational wave
losses, and given the model assumptions (BH spin effects). Therefore, a
quantitatively convincing assessment cannot be presented for this
possibility. 

\subsection{Dynamical evolution}

We describe the merger for the two most extreme cases 
(non-spinning BH) in our set of models: 
(1) a small BH with large NS, specifically a mass ratio of $Q\!=\!2$
and a small value for compactness of ${\cal C}=0.145$; 
(2) a large BH with a small NS, specifically a mass ratio $Q\!=\!5$
and a value for compactness of ${\cal C}=0.180$. 
These two models cover the main features seen in the simulations,
while other models with intermediate values for $Q$ and ${\cal C}$ are
easily recognised as being similar to one or the other of the two cases.

\subsubsection{Low mass ratio, small compactness}

\begin{figure*}
   \centering
   \begin{tabular}{cc}
   \includegraphics[width=8.5cm]{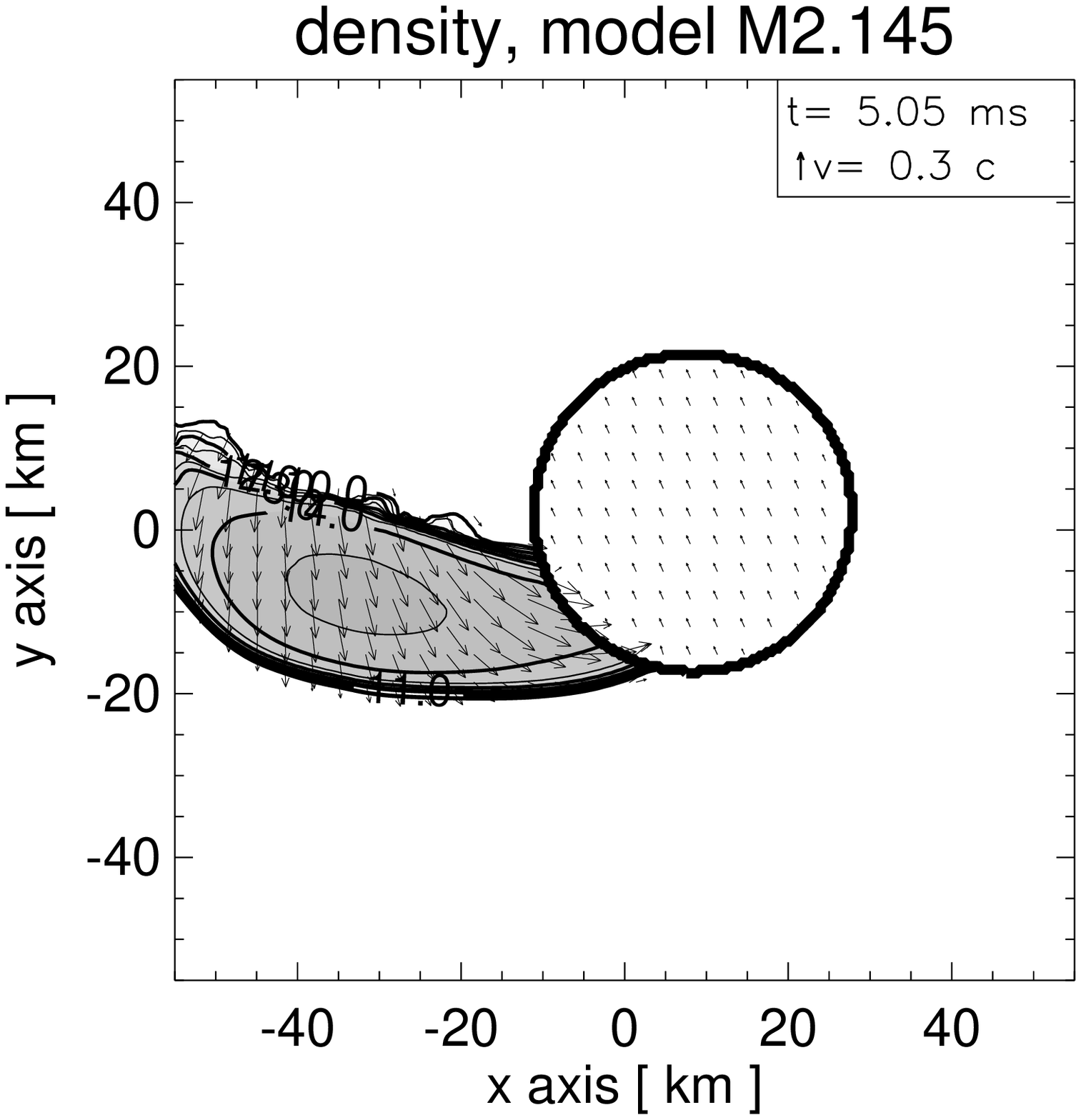} &
   \includegraphics[width=8.5cm]{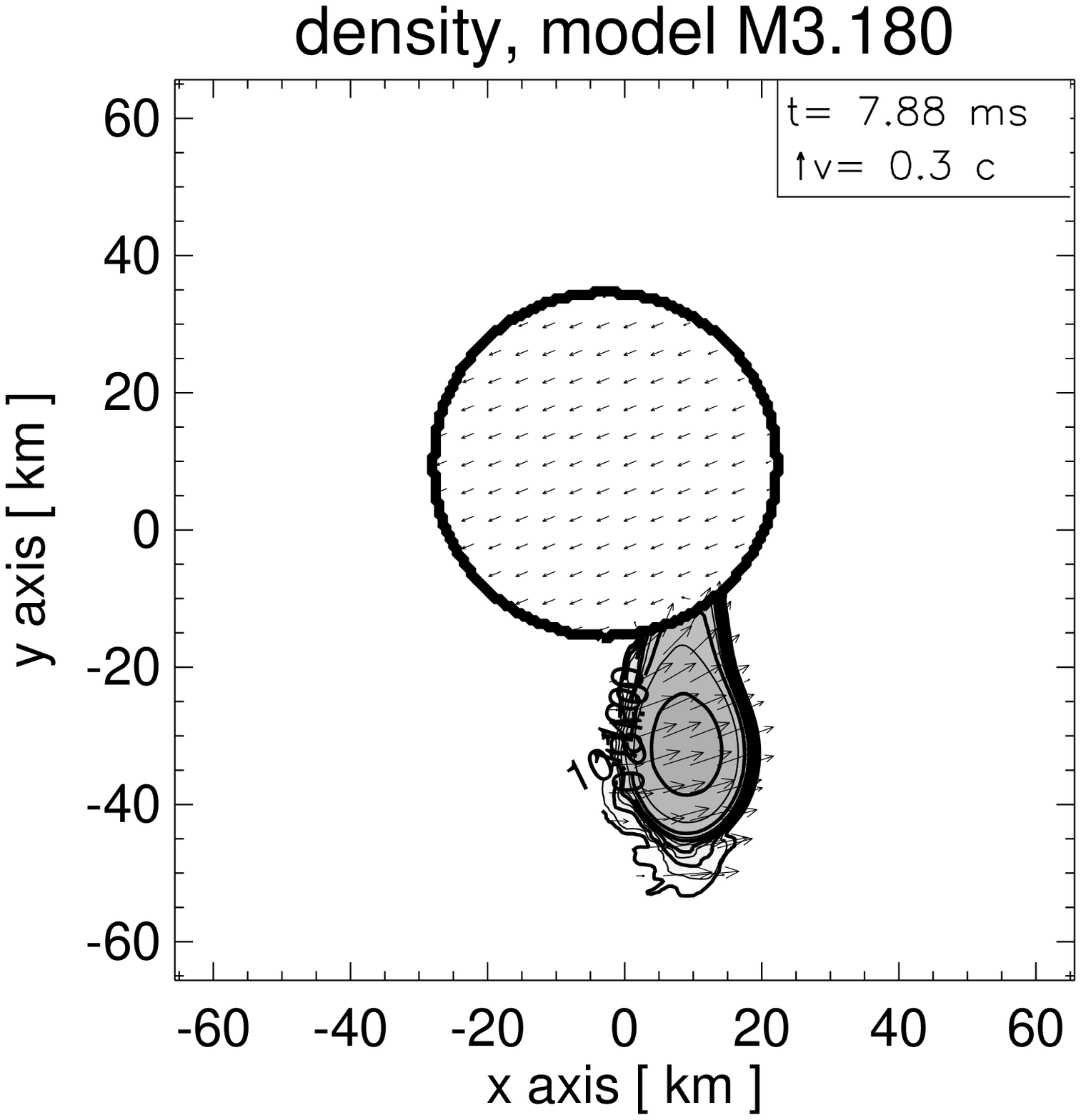} \\[-0.5ex]
   \includegraphics[width=8.5cm]{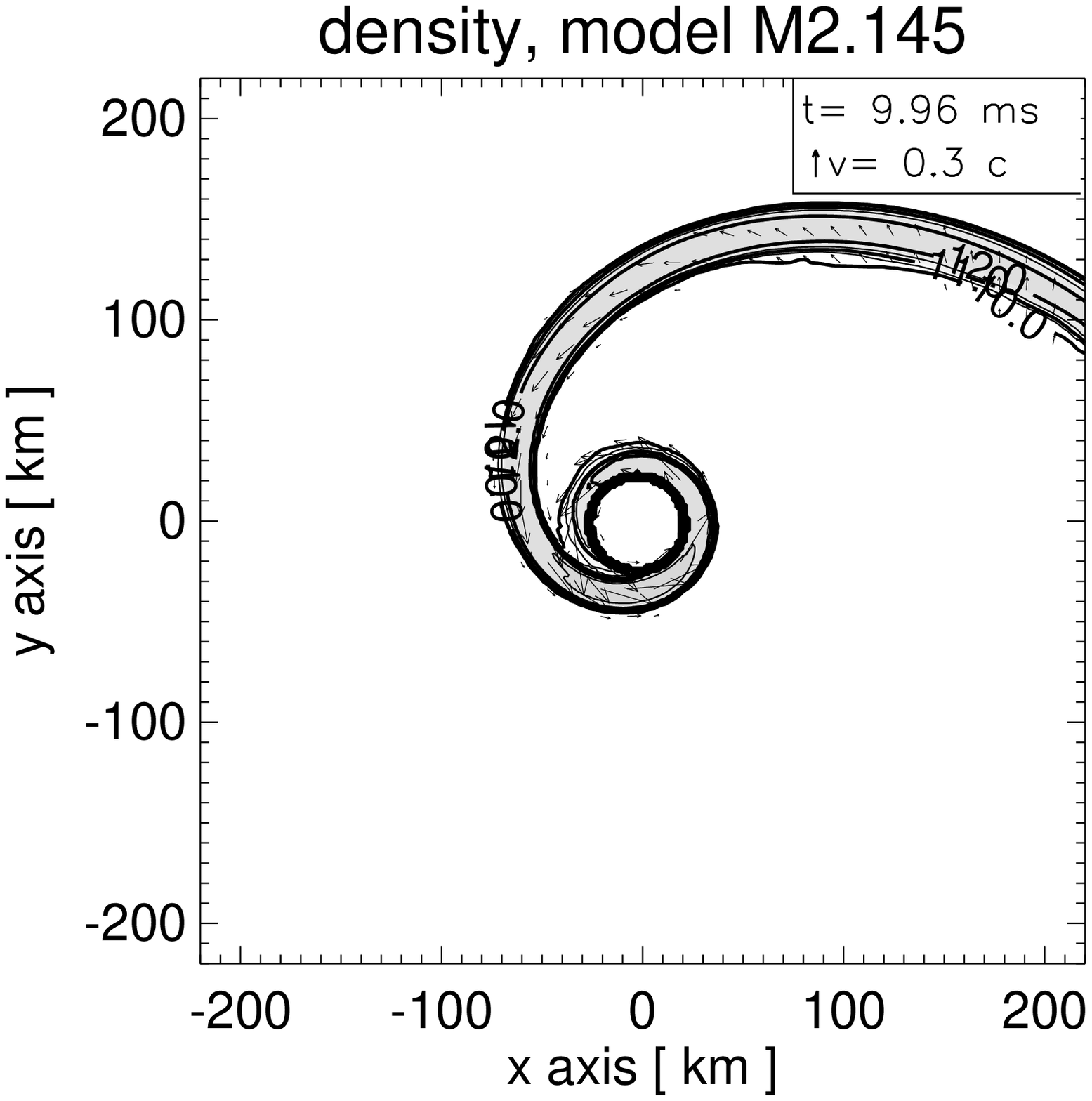} & 
   \includegraphics[width=8.5cm]{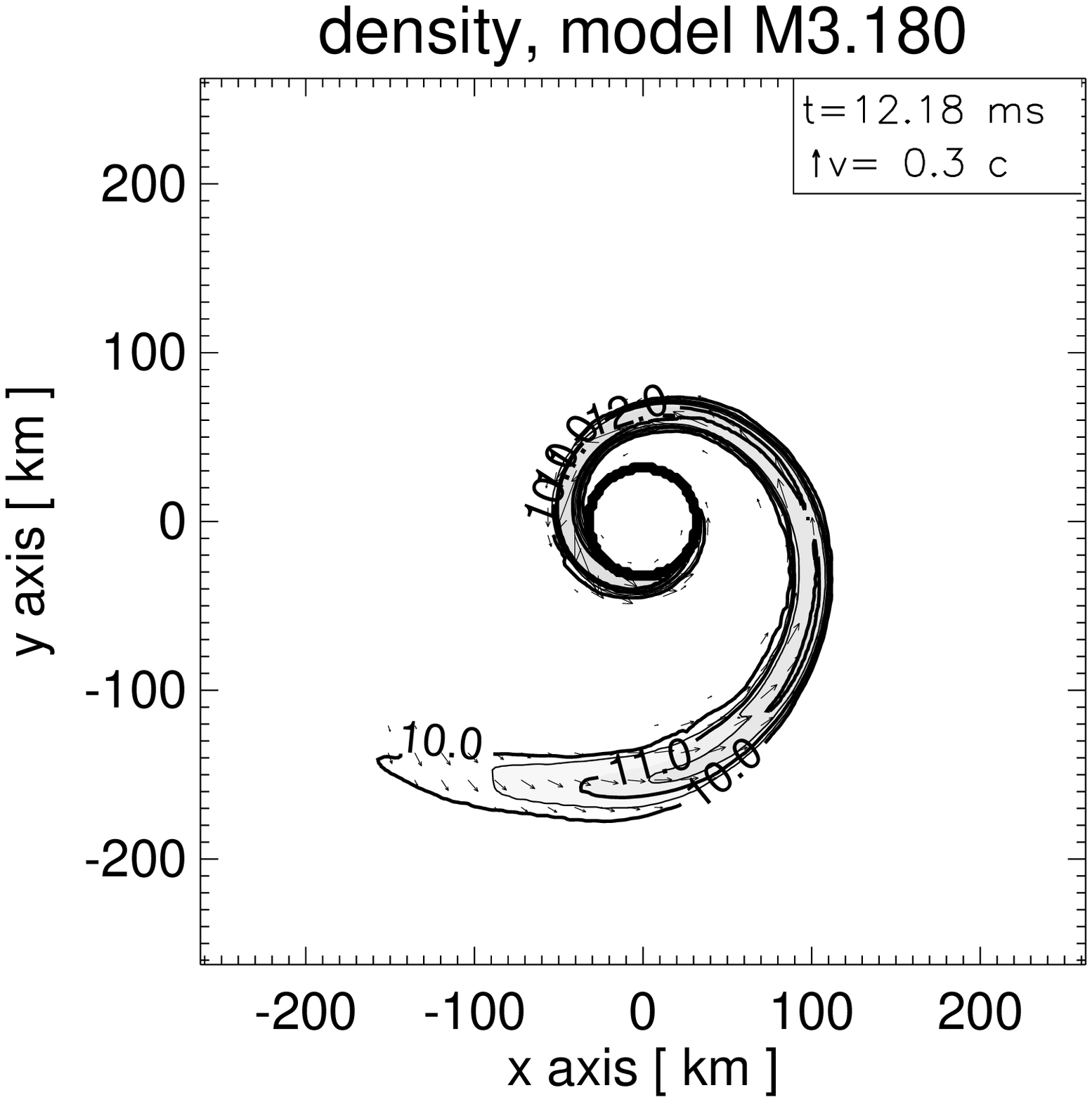} \\[-0.5ex]
   \includegraphics[width=8.5cm]{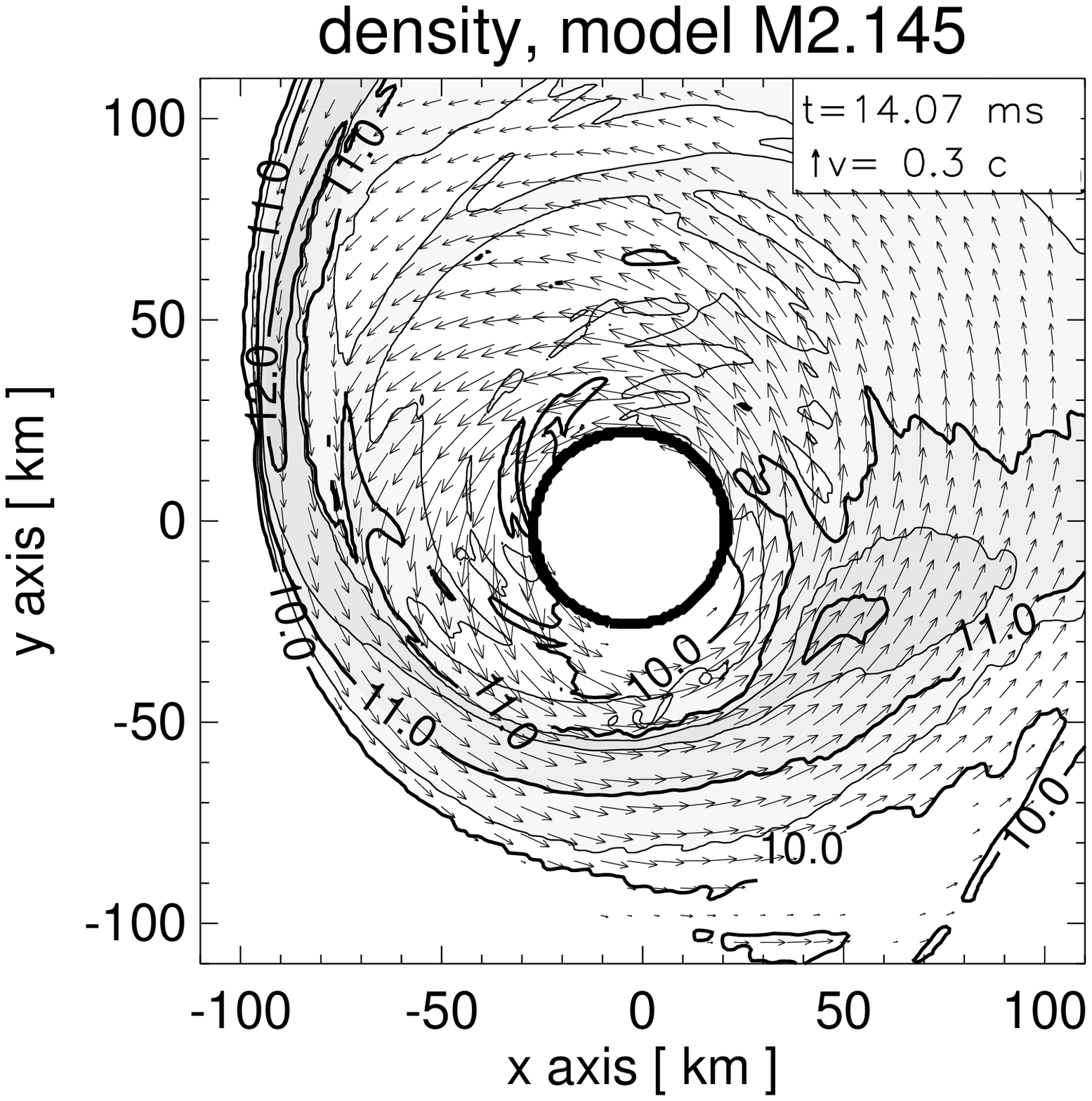} &
   \includegraphics[width=8.5cm]{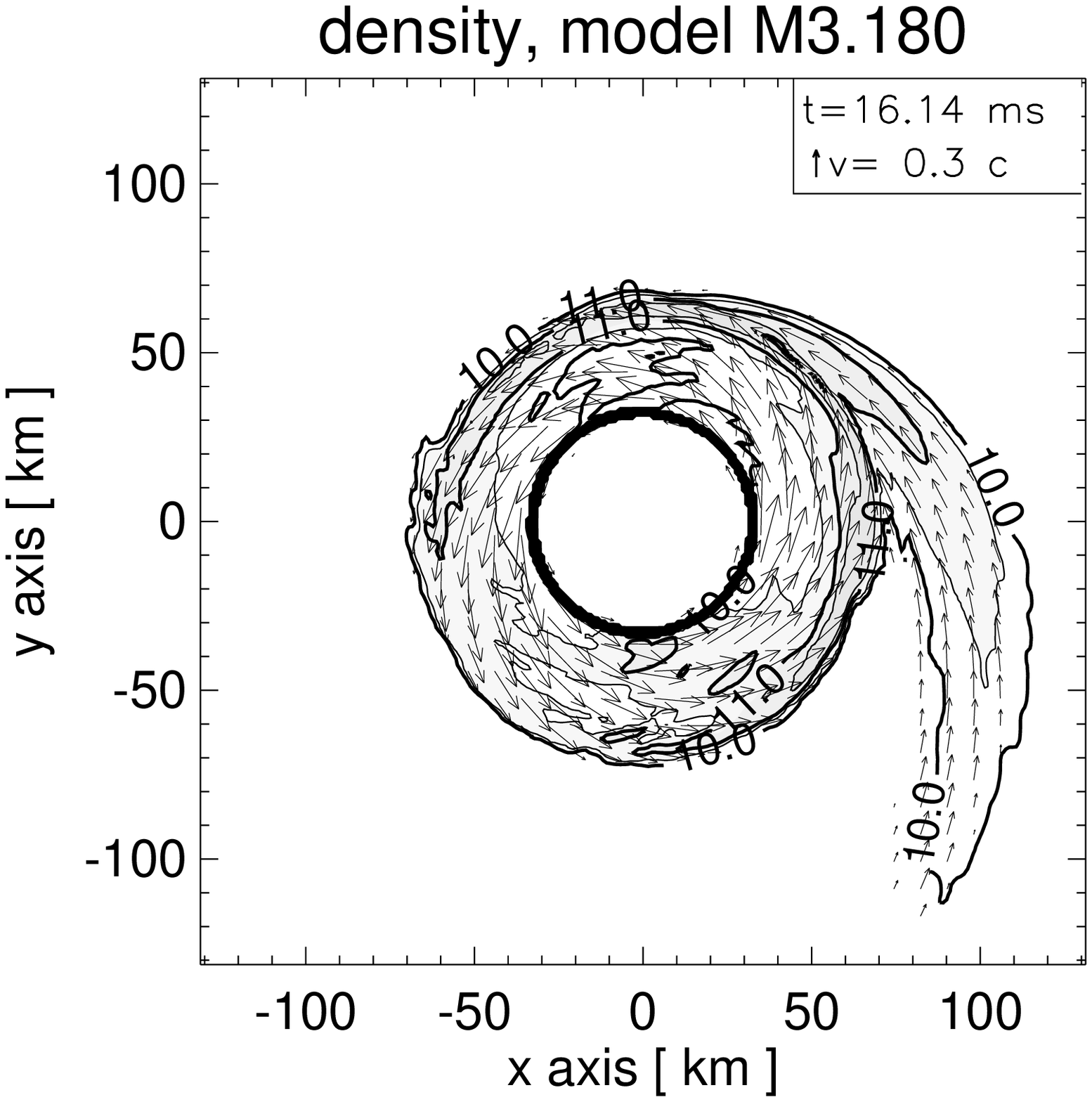} \\[-2ex] 
   \end{tabular}
   \caption{The mass density distributions for models M2.145 and M3.180
     are displayed in the orbital plane with contours spaced
     logarithmically in steps of 0.5 dex, 
     with units g$\,$cm$^{-3}$. The arrows indicate the velocity field.
     The circle at the centre outlines the BH radius, which is the
     arithmetic mean of the event horizon and ISCO. }
   \label{fig:contour1}
\end{figure*}

During the initial orbital decay of model {\em M2.145}, the central
density of the NS remains fairly constant (Figs.~\ref{fig:densi} and
\ref{fig:mass}). 
One notices small oscillations originating in the initially spherical NS
adapting to the potential of the BH and the discretisation on the grid. 
At about 5~ms, the closest distance of the NS density maximum or
centre-of-mass is reached (Fig.~\ref{fig:sep} bottom left;
Fig.~\ref{fig:contour1} top left).  
At this point, matter is transferred to the BH, but at a
relatively slow rate compared to the other case. 
The NS is tidally disrupted, but not before a substantial portion
of its matter is given an elliptic orbit taking it temporarily away from
the BH: we note the increase in distance of the density maximum in all
cases of low mass ratio and small compactness (models on the lower
and left hand side in Fig.~\ref{fig:sep}). 
We return to this point of orbital widening in Sect.~\ref{sect:orbit}.
The ``radius of the BH'' shown in Fig.~\ref{fig:sep} is the 
arithmetic mean of the event horizon and innermost stable circular
orbit, as described in Sect.~\ref{sect:initcond}. 
For a non-rotating BH, this is equal to two Schwarzschild radii.

The NS matter spreads out in a tidal spiral arc containing 
$0.08\,M_\odot$ (Fig.~\ref{fig:contour1}, central left panel 
$t\!=\!9.96\,$ms).
Eventually matter falls back along the spiral arc toward the BH. 
Some of it manages to complete a full circle around the BH and then
collides with the incoming material. 
At this point, the infalling matter is deflected away from the BH, the
circling material spreads out to fill the space and a torus is formed. 
The point in time when this deflection occurs is marked by a short
vertical line in Fig.~\ref{fig:mass}, and one can see that this
applies only to models with low mass ratios of $Q\!=\!2$ and $Q\!=\!3$, 
for all three values of compactness.  
In some cases, the high-resolution models were not evolved
sufficiently long to see the formation of a disk. 

A fraction of the material in the spiral arc is unbound. 
This amount is also indicated in Fig.~\ref{fig:mass}, by a short piece
of near-horizontal line. 
The values are listed in Table~\ref{tab:models}.  

\subsubsection{High mass ratio, large compactness}

During the initial orbital decay of model {\em M5.180}, the central
density increases substantially (Fig.~\ref{fig:densi}, similarly to
model M4.180) in the calculations of sufficiently high resolution. 
This increase can be noted for all models with massive BHs, i.e., high
mass ratios (from close inspection of Fig.~\ref{fig:mass}). 
So when the NS is close enough to finally transfer mass to the BH, it
is very compact and mass transfer proceeds relatively quickly: we note
the very steep decline of the mass on the grid for this model M5.180,
in contrast to M2.145 described above. 
This effect is also partially caused by an increase of the mass and
radius providing an essentially direct approach to the absorbing
surface for large BHs.

\begin{figure*}
   \centering
   \begin{tabular}{cc}
   \includegraphics[width=8.5cm]{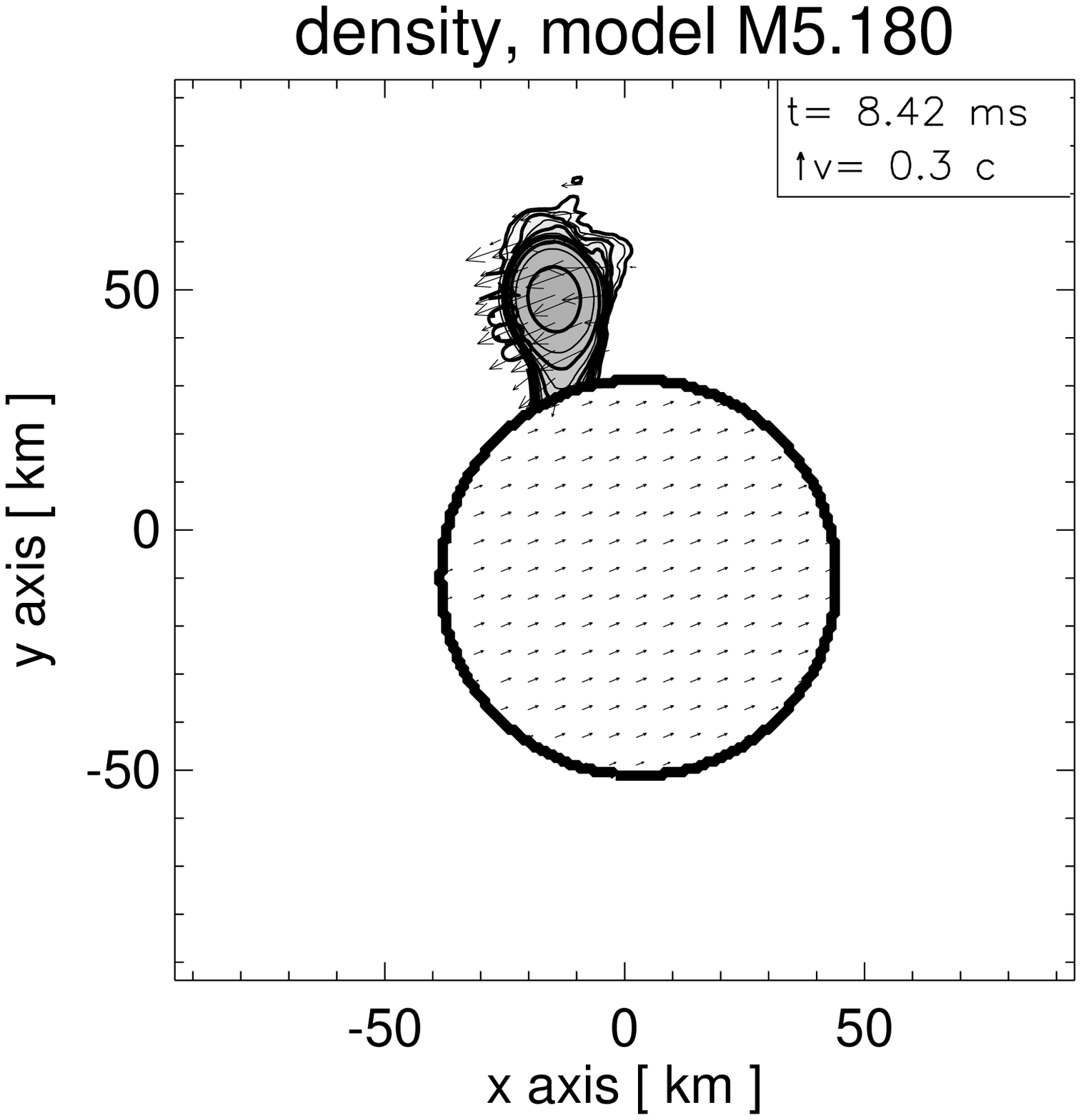} &
   \includegraphics[width=8.5cm]{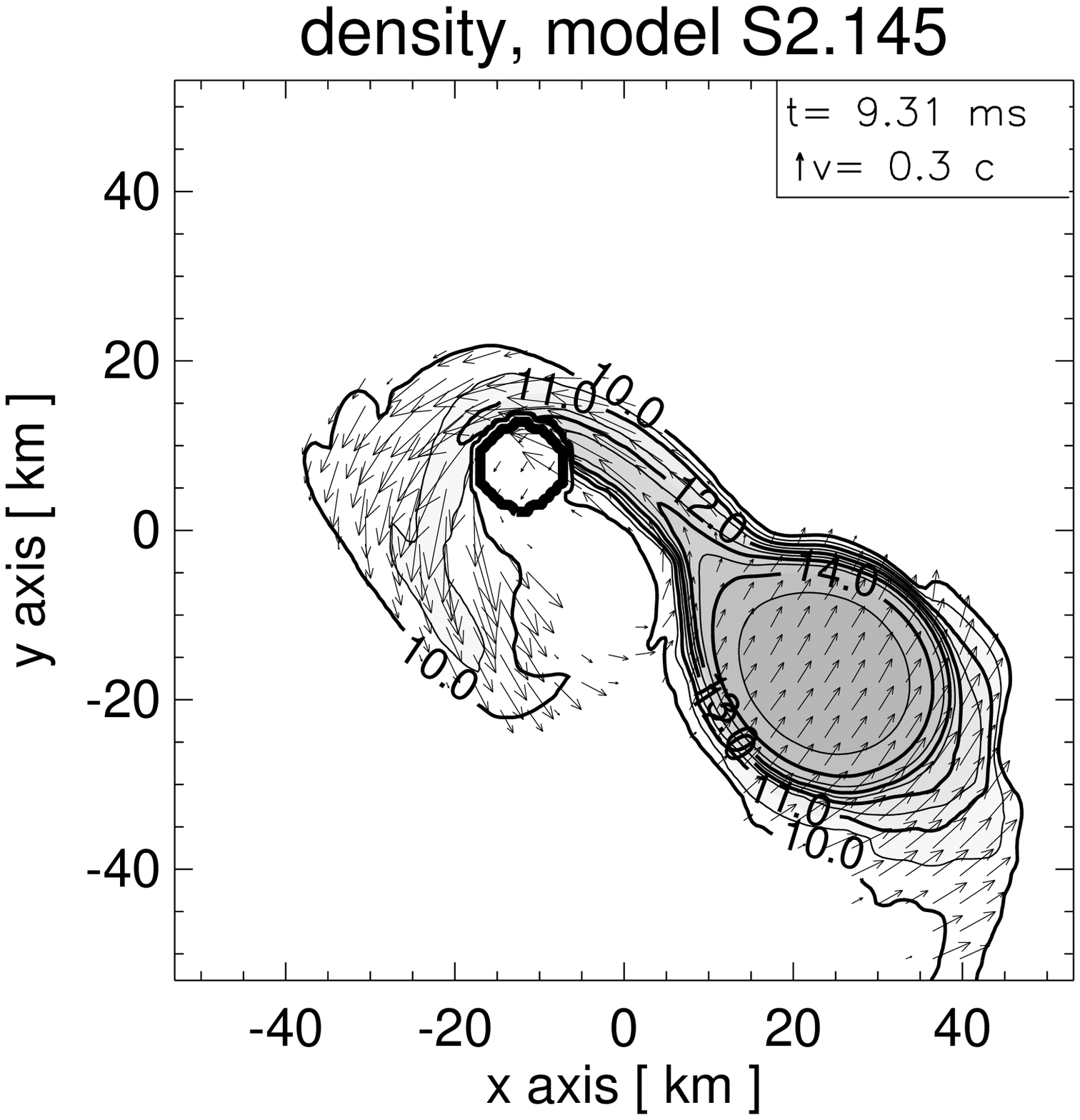} \\[-0.5ex]
   \includegraphics[width=8.5cm]{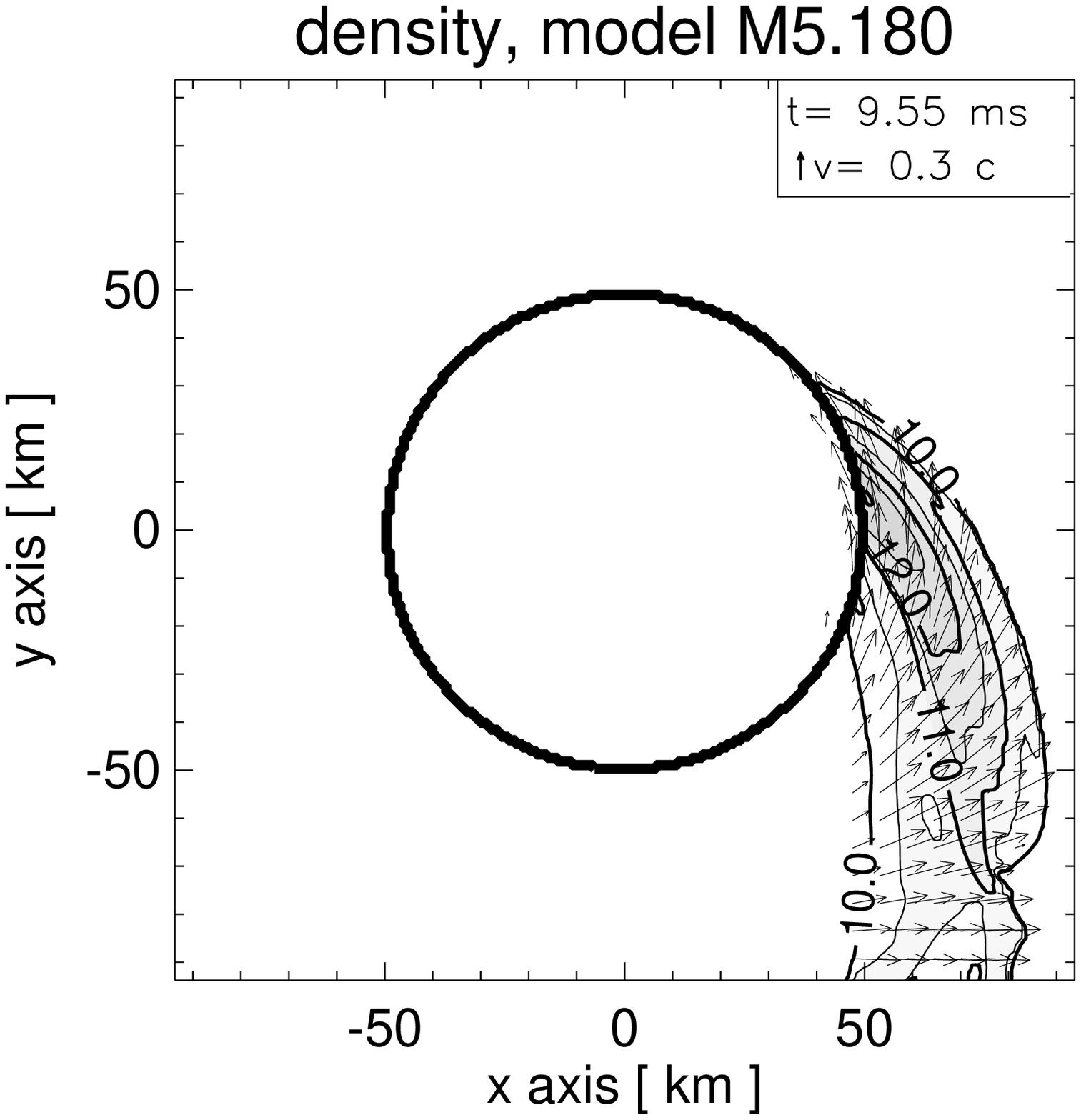} &
   \includegraphics[width=8.5cm]{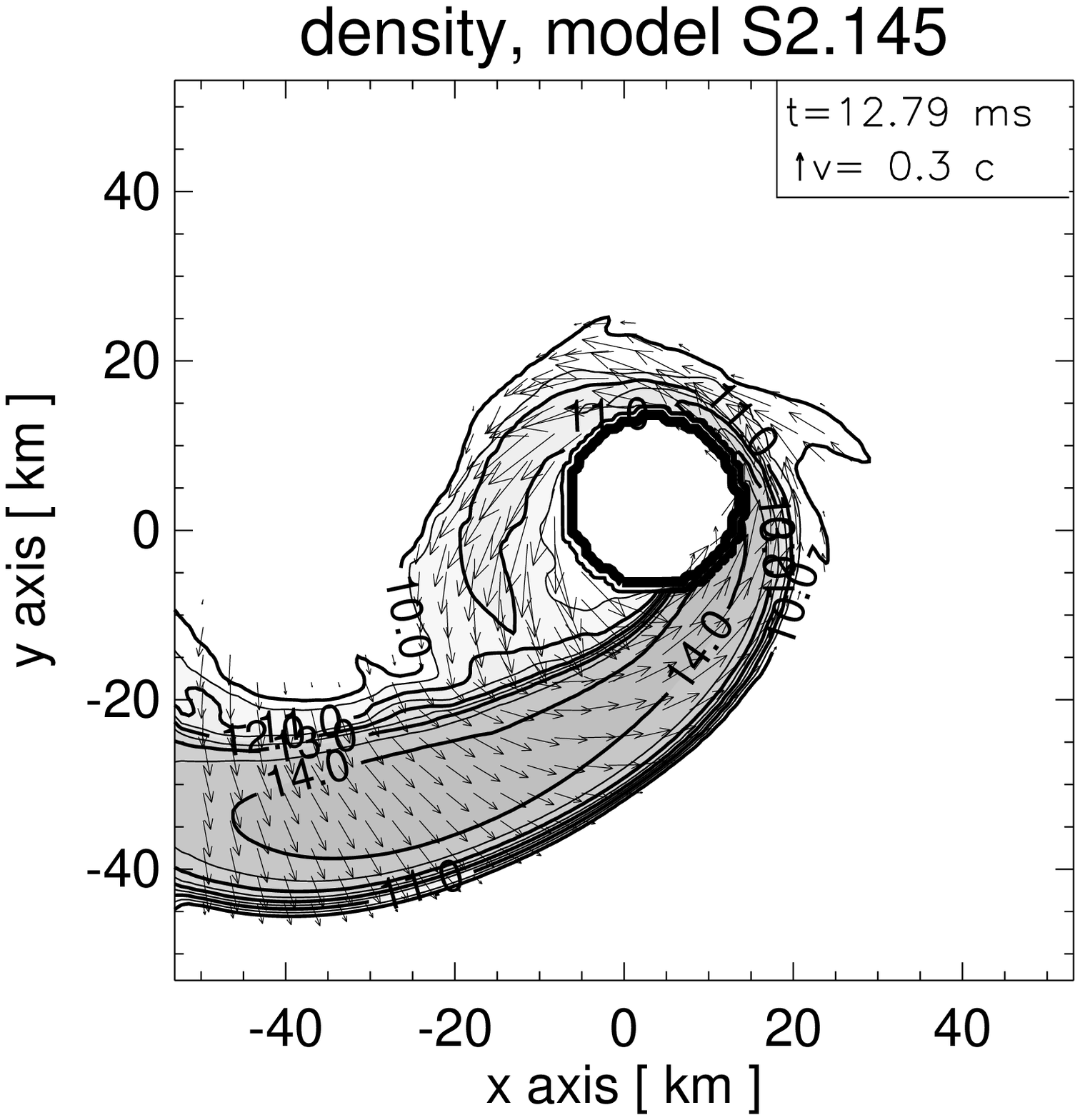} \\[-0.5ex] 
   \includegraphics[width=8.5cm]{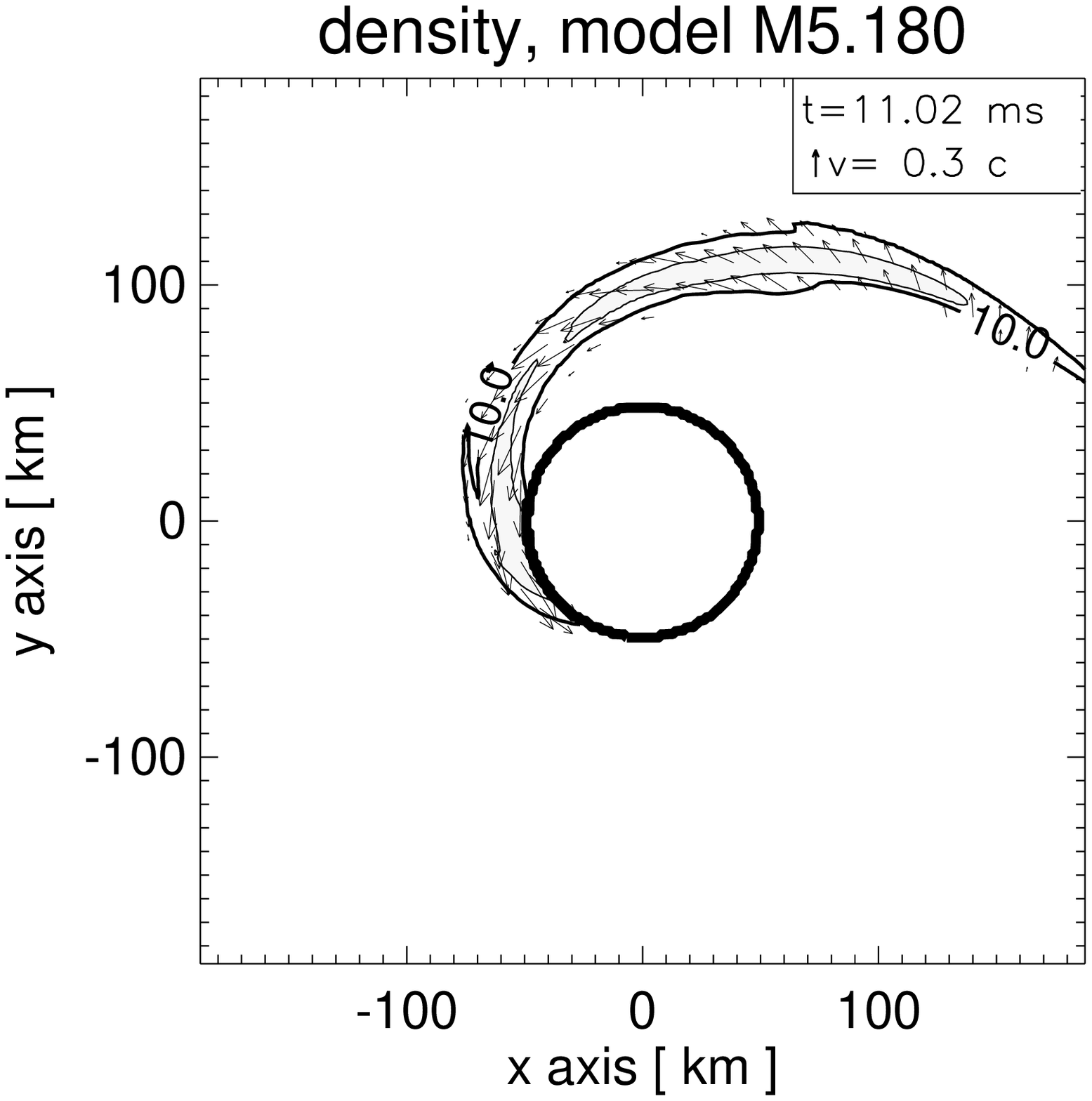} &
   \includegraphics[width=8.5cm]{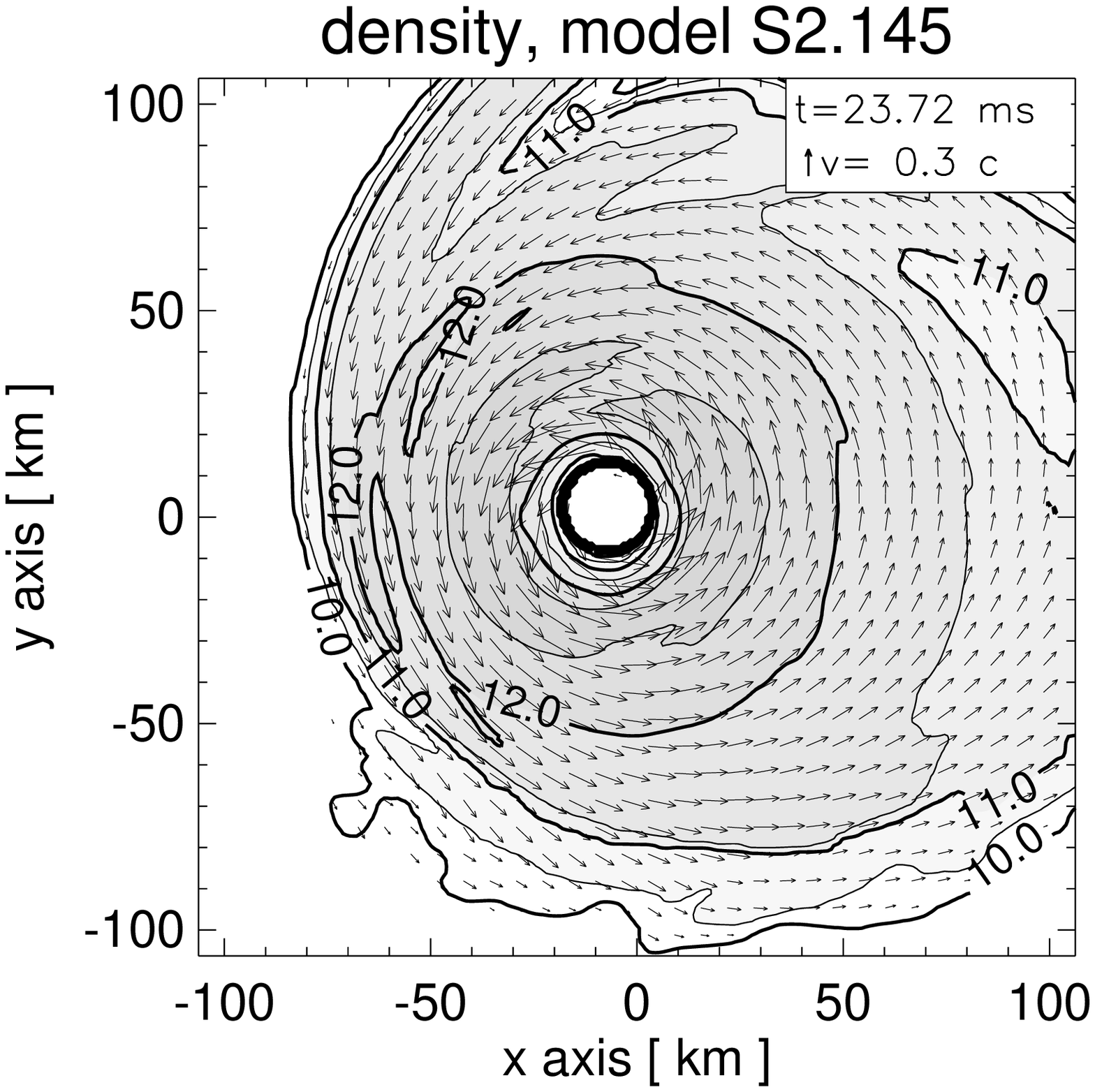} \\[-2ex] 
   \end{tabular}
   \caption{The mass density distributions for models M5.180 and
     S2.145 are displayed in the orbital plane with contours spaced
     logarithmically in steps of 0.5 dex, with units g$\,$cm$^{-3}$. 
     The arrows indicate the velocity field.
     The circle at the centre outlines the BH radius, which is the
     arithmetic mean of the event horizon and ISCO. }
   \label{fig:contour2}
\end{figure*}

The distance from the BH at which NS density reaches a maximum
decreases continuously all the way to the numerical surface of the BH. 
In the right panels of Fig.~\ref{fig:sep}, one can see the line of
separation that nearly meets the line of BH radius to within a few km. 
The minimal distance shown in the opposite case of model M2.145 is
much larger, approximately 15~km.   

The NS matter remaining in model M5.180 after the merger, also spreads
out in a tidal spiral arc but contains only $10^{-3}\,M_\odot$ 
(Figs.~\ref{fig:mass} and~\ref{fig:contour2}).
The BH, which has a very large extent, accretes all the material that
flows back along the arc; none manages to flow all the way around the
BH and form a disk. 

Eventually the matter in the outermost portions of the spiral arc
reaches the edge of the largest grid (only the lower resolution
simulations can be evolved for this length of time).
For most models, this happens at around 15-20~ms (depending on the
model) at which point the amount of unbound matter ejected from the
system is registered by the calculation. 
The amount of matter ejected is approximately
$10^{-3}$--$10^{-2}$~$M_\odot$. 
We also calculate the amount of material on the grid that is not
gravitationally bound, by comparing the internal and kinetic
energies with the potential. 
The values of these mass components are listed in
Table~\ref{tab:models} and are consistent with the amounts that
actually flow off the coarsest grid as described above. 
Dividing the distance travelled by this material -- roughly half the
grid size $L/2\!\approx\!400$~km -- by the time elapsed between merger
and arriving at the edge of the grid (approximately 5~ms), yields an
average speed of c/4.

\begin{table*}
\caption{Key initial model parameters and some results for models
  including BH spin.}  
\label{tab:Rot}      
\centering          
\begin{tabular}{c c c c c c c c r c c c c c} 
\hline\hline       
Model & $Q$ & $\cal{C}$ & $M_{\mathrm{BH}}$ & $a_{\mathrm{i}}$ & 
$R_{\mathrm{NS}}$ & $d_{\mathrm{i}}$ & $N$ & 
$L$ & $\Delta x$ & $M_{\mathrm{u}}$  & $M_{\mathrm{b}}$ & 
$M_{\mathrm{g}}$ & $a_{\mathrm{f}}$   \\ 
      &     &           & $M_\odot$   &                     &
 km                   & km        &         &
 km  &  km       & $M_\odot$  &          $M_\odot$ & 
  $M_\odot$  &   km    \\ 
\hline
 R5.180 & 5 & 0.180 & 7.0 & 0.00 & 11.5 & 88 & 256 & 1450 & 0.71 & 0.11 & 0.07
  & 0.18 & 0.39  \\
 R3.160 & 3 & 0.160 & 4.2 & 0.00 & 14.3 & 66 & 256 & 1100 & 0.54 & 0.08 & 0.14
  & 0.22 & 0.44  \\
 R2.145 & 2 & 0.145 & 2.8 & 0.00 & 14.3 & 56 & 128 &  880 & 0.86 & 0.05 & 0.17
  & 0.22 & 0.56  \\
 S2.145 & 2 & 0.145 & 2.8 & 0.99 & 14.3 & 52 & 128 &  850 & 0.83 & 0.02 & 0.26
  & 0.28 & 0.88  \\
\hline                  
\end{tabular}
\begin{list}{}{}
\item[] Initial
  parameters: mass ratio $Q$, compactness $\cal{C}$, black hole mass
  $M_{\mathrm{BH}}$, initial BH spin parameter $a_{\mathrm{i}}$, 
  neutron star radius $R_{\mathrm{NS}}$, initial orbital
  distance $d_{\mathrm{i}}$, number of zones per dimension $N$, size
  of largest grid $L$, size of finest zone $\Delta x = L/8N$. 
  The mass of the NS is $1.4\,M_\odot$ in all cases.
  Values at the end of the simulation: 
  unbound+ejected gas mass $M_{\mathrm{u}}$, 
  bound neutron star mass around BH $M_{\mathrm{b}}$, 
  neutron star mass not instantly accreted by BH $M_{\mathrm{g}}$,
  final BH spin parameter $a_{\mathrm{f}}$. 
\end{list}
\end{table*}

\subsubsection{Comparison of models}

Looking at Figs.~\ref{fig:mass} and~\ref{fig:sep}, one sees that all
models with ${\cal C}=0.145$ and ${\cal C}=0.160$  
follow the patterns described above for model {\em M2.145}. 
All models with ${\cal C}=0.180$ follow the mass accretion pattern of 
{\em M5.180}, although some aspects of models M2.180 and M3.180 are
different: the BH is small enough in these two cases to allow
formation of a disk, albeit with a very small mass. 
We included the information about the mass remaining on the grids in
Fig.~\ref{fig:ISCO.MS} in an attempt to outline the similarities
between the various models.

\subsection{Effects of BH spin\label{sect:SpinBH}}

We calculated four models to assess the influence of BH spin:
in three models (R2.145, R3.160, R5.180), the BH is taken to be
initially non-rotating ($a_{\mathrm{i}}\!=\!0$), but we allowed a
change of the BH potential due to spin-up by accretion of angular
momentum of the NS matter during the evolution. 
The opposite case, an initially rapidly spinning BH with
$a_{\mathrm{i}}=0.99$,  
is simulated in model S2.145, and like the other three models,
the spin and potential are allowed to change by accretion as well. 
The complete investigation of the effects of these changes is
outwith the remit of this paper and is left to future work. 
Here we only point out the boundaries of the relevance of 
results obtained with models ignoring BH spin. 

\begin{figure*}
 \centering
 \begin{tabular}{cc}
   \includegraphics[bb=30 110 390 755,clip,width=13.7cm]{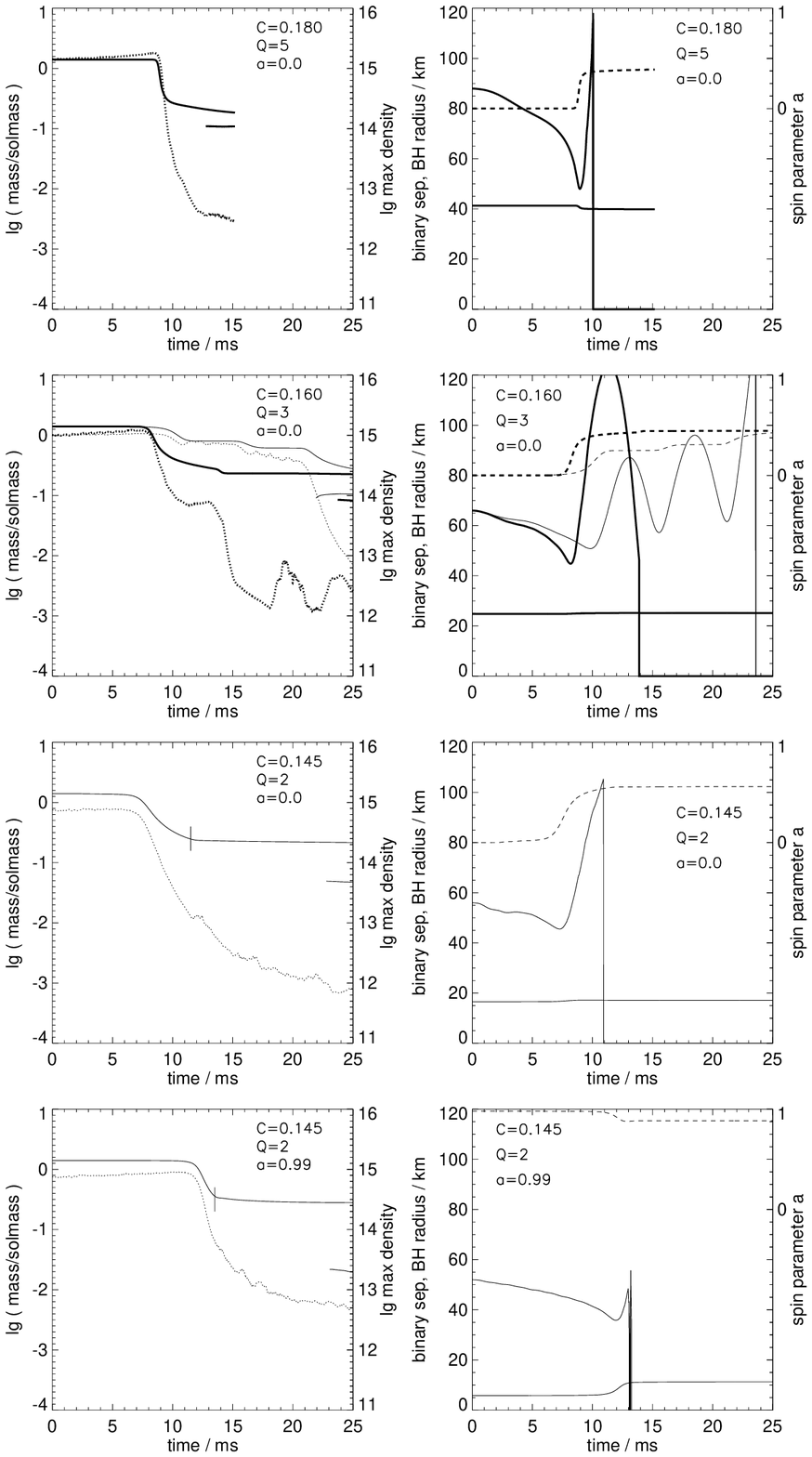} &
       \begin{minipage}[b]{3.8cm} \caption{ \sloppy
     Temporal evolution for models R5.180, R3.160, R2.145, and
     S2.145 (from top to bottom), which include the effect of spin-up
     of the BH (see Table~\ref{tab:Rot}). The initial value of the BH
     rotation parameter is given in the panels. 
     Thin lines show lower-resolution simulations ($128^3$);
     bold lines show higher-resolution simulations ($256^3$).
     The {\em left column} of panels shows: the gas mass (solid lines) and
     the maximum density (dotted lines). A short vertical line
     indicates the formation of an accretion disk. Short pieces of
     nearly horizontal lines show the amount of unbound mass.
     The {\em right column} of panels shows: the radius of the BH (arithmetic
     mean of horizon and ISCO) as lower solid line, the binary
     separation as upper solid line, and the BH spin parameter $a$ as
     top dashed line. 
     The vertical line indicates when the mass within a sphere around
     the density maximum falls below $0.03\;M_\odot$. }
    \label{fig:rots0}
   \end{minipage}
  \end{tabular}
\end{figure*}

The post-merger rotational speed of the BH {\em in}creases monotonically
within the three models with initial $a_{\mathrm{i}}\!=\!0$ spin, from
a value of 
$a_{\mathrm{f}}\!=\!0.39$ to $a_{\mathrm{f}}\!=\!0.56$ (see
Table~\ref{tab:Rot} and Fig.~\ref{fig:rots0}) when {\em de}creasing
the mass ratio and compactness. 
Following the prescription of Eq.~(\ref{eq:ArtePaWi}), an increasing
spin reduces both the radius of the horizon and the radius of the ISCO
(Fig.~\ref{fig:rots0}).  
In these models, the decrease in radius caused by spin is greater than
the increase caused by mass, or they balance it out, so one observes a
slight overall decrease in effective radius (model R5.180) or hardly
any change (models R3.160, R2.145): we note the lower near-horizontal
line in the right panels of Fig.~\ref{fig:rots0}.

If at the closest point when the NS material is being accreted by the
BH, the radius of the BH remains the same or even shrinks --- and with
it the effective potential --- the effectiveness of matter accretion
is reduced. 
In the end, after tidal disruption, much more material remains in the 
spiral arc than in the equivalent models without BH spin, e.g.,
compare R5.180 with M5.180.
When the models that include the effect of BH spin are completed, 
between $0.18\,M_\odot$ and $0.28\,M_\odot$ of NS material
is dispersed around the BH (see Table~\ref{tab:Rot} and 
Sect.~\ref{sect:dmass} for further discussion).

The initially rapidly rotating BH (model S2.145) is decelerated
from $a_{\mathrm{i}}\!=\!0.99$ to $a_{\mathrm{f}}\!=\!0.88$
(Table~\ref{tab:Rot} and Fig.~\ref{fig:rots0})
because the initial BH with $a_{\mathrm{i}}\!=\!0.99$ 
has a higher specific angular momentum than the matter that is
subsequently accreted.
In addition to the increase in mass, the reduction in rotation itself
increases the size of the BH: we note the large increase in the
circle radius in the right panels in Fig.~\ref{fig:contour2} and the
rise of lower solid line in the bottom right panel of Fig.~\ref{fig:rots0}.  

Model R3.160 is the only one presented in this paper that exhibits a
distinctive orbit widening, the lower resolution model even showing
repeated mass accretion episodes.  
In this case, the NS remains self-bound, albeit being less massive,
after the first close encounter with the BH. 
In the high-resolution simulation the NS is tidally disrupted during
the second encounter, while for the low-resolution simulation it takes
another two episodes before the NS is tidally disrupted. 
The orbital distance plot (Fig.~\ref{fig:rots0}, right panel, second
from top) resembles those from our own previous works 
(Janka et al.~1999) and others (e.g., Davies et al.~2005).  
We return to this point below.


\section{Discussion\label{sect:discuss}}

We discuss three specific points: (1) the orbit widening during the
accretion and tidal disruption of the NS, (2) the amount of 
mass remaining to produce a disk, and (3) the consequences of the NS
reaching the ISCO before being tidally disrupted. 

\subsection{Orbit widening\label{sect:orbit}}

In previous simulations that adopted a purely Newtonian potential for
the BH, it was noted that during mass accretion from the NS into the BH,
the radial position of the density maximum shifted outwards. 
This resulted in a lighter NS moving away from the BH on an elliptic
orbit. 
Of course, the shape of the orbit and the gravitational wave emission
eventually caused the NS to experience a second interaction with the BH,
in the meantime the accretion, however, being effectively shut off. 
The sequence of mass transfer and orbit widening can be repeated. 
Numerically this was observed e.g., in Lee \& Klu\'zniak~(1999; their
Fig.~20 shows the binary separation), Janka et al.~(1999; their Fig.~1
shows the orbital separation) and in Rosswog et al.~(2004; their
Fig.~7 shows the mass accretion episodes).
It was explained with a semi-analytic description by Davies et
al.~(2005), in which some ``fraction of the angular momentum of the
matter (transferred from the NS to the BH) is fed back into the NS''.
Their Fig.~1 shows the orbital separation.

Miller (2005) argued that the presence of an ISCO, which is a purely
relativistic effect, would substantially alter the outcome of a NS--BH
merger, since the NS would be preferentially absorbed rather than
tidally disrupted. This is basically corroborated by in the general
relativistic simulations of Rantsiou et al.~(2008), Etienne et 
al.~(2009) and Shibata et al.~(2009; their Fig.~1 shows the orbital
separation without any sign of orbit widening).  

The Paczy\'nski-Wiita potential in our simulations mimics the ISCO, so
the effect outlined by Miller (2005) should be present in our
otherwise Newtonian simulations. 
Indeed, the plots of the distance between the centre of the BH and the
centre-of-mass of the NS (Fig.~\ref{fig:sep}, Fig.~\ref{fig:rots0})
place the results of our simulations squarely in-between the Newtonian
and the relativistic results: we observe a beginning of an orbit
widening, but the widening is not fast enough to produce a self-bound
`mini'-NS.   
In particular, in the models with high mass ratios and highly compact
NSs, the binary distance shrinks to the `surface' of the BH and
hardly any subsequent orbit widening is seen. 
The NS is practically absorbed whole, as noted in the relativistic
calculations. 

An exception is model R3.160 (including the potential changing effects
of BH spin), which exhibits the repeated mass accretion episodes that
are very similar to the purely Newtonian models (see
Fig.~\ref{fig:rots0} right panel second from top).
This might be the consequence of two opposing effects. Firstly, the orbit
widening is more pronounced in models that include BH spin (compare 
right panels of Fig.~\ref{fig:rots0} with the equivalent models in
Fig.~\ref{fig:sep}, because the denser regions of the `escaping' NS
are caught up to a lesser degree within the smaller BH
radius. Secondly, model R3.160 is 
intermediate between R5.180 where the large BH swallows up the compact
NS, and model R2.145 where a small BH produces a large differential
tidal field on an extended NS. 
In this latter case, the NS is tidally disrupted, although its orbit
has widened significantly. 
The remaining mass in model R3.160 manages to recollect itself into a
roughly spherical NS on an elliptic orbit. 
We finally note, however, that the number of mass-transfer episodes
of this model still varies between the simulations of different
resolution, so the detailed values and sequence of events (e.g., number
of cycles) have to be interpreted with caution.

\subsection{Disk mass\label{sect:dmass}}

The amount of material not `instantly' absorbed by the BH but remaining
in the surroundings and potentially available to form a torus, is an
important energy source for powering the jet of a gamma-ray burst. 
Thus the questions of how much matter is accreted, how much is
ejected, and how much forms a torus, need to be investigated
carefully. 
In this point, the numerical general relativistic simulations 
(Rantsiou et al.~2008; Etienne et al.~2009; 
Shibata et al.~2009, their Fig.~7) still differ markedly among
themselves, even for models with similar parameters. 
The range spans from $\approx\!0.1\,M_\odot$ to less than
$10^{-4}\,M_\odot$. 
To power gamma-ray bursts effectively, at least some hundredths of a
solar mass appear to be needed.  

The values of mass remaining around the BH for our models are tabulated
in Tables~\ref{tab:models} and~\ref{tab:Rot}.
Only in models with both a compact NS (${\cal C}\!=\!0.180$) and a
non-rotating BH, does most of the matter become accreted quickly by
the BH and only $10^{-2}$--$10^{-3}$~$M_\odot$ remain to form a torus,
especially for the massive BHs with a high mass ratio $Q$. 
This trend matches the general relativistic findings; 
however, the absolute quantities of mass are higher in our models,
e.g., Shibata et al.~(2009) obtain typically 1\% of the neutron star
mass for models of low  mass ratio $Q$.
On the other hand, Etienne et al.~(2009) infer $0.06\,M_\odot$ for a
non-rotating BH model with $Q\!=\!3$. 
For less compact NSs, we obtain around $0.1\,M_\odot$.   

As already mentioned above, our models including the effect of BH spin
(Table~\ref{tab:Rot}) end with a significant amount (between
$0.18\,M_\odot$ and $0.28\,M_\odot$) of NS material spread out around
the BH, as the BH's effective potential depth decreases because the
spin-up matches or outweighs the mass increase (Fig.~\ref{fig:rots0},
models R5.180, R2.145). 
 
\subsection{Tidal disruption vs.~ISCO}

Following Miller's (2005) argument, we would expect little mass 
(tenth of a percent of a solar mass, or less) to remain around the BH
if the NS reaches the ISCO before being tidally shredded. 
On the other hand, if the NS were tidally disrupted (`mass shedding',
MS) before reaching the ISCO, then significant amounts of matter
(several percent of a solar mass) would be able to form a torus around
the BH. 
With this in mind, and to be able to compare with previous relativistic
simulations, we chose the model parameters plotted in
Fig.~\ref{fig:ISCO.MS}. 
This graph also uses different symbols to show the varying remaining
gas mass around the BH. 
No clear correlation pattern can be seen between the remaining gas
mass and the expected line separating the ISCO from the MS case,
except a general trend that the amount of mass is indeed lower toward
the right (models with higher compactness). 
However, we cannot identify a clear step change. 
This lack of clear demarcation between ISCO and MS is also the case for
the models including general relativity (Shibata et al.~2009,
Fig.~\ref{fig:ISCO.MS.Shi}). 

One point to remember is that the mass ratio $Q$ and compactness
${\cal C}$ used in this paper, are based on the purely Newtonian,
uniquely defined, masses of NS and BH, and radius of the NS. 
Of course, in the models with general relativistic physics, $Q$ and
${\cal C}$ have to be defined with a specific choice of the ADM mass,
rest mass, circumferential radius, isotropic radius, etc. 
This complicates the comparison between our Newtonian results shown in
Fig.~\ref{fig:ISCO.MS} and the general relativistic work of others
in Fig.~\ref{fig:ISCO.MS.Shi}. 
We would expect these differences to produce a shift in the
separating line within the plot, but guess it would only be by several
tens of percent, and therefore would not qualitatively change the
statements. 

\subsection{Rotating BH}

For models R5.180, R3.160, R2.145, and S2.145, we note only that a
large difference in dynamics and remaining torus mass is present
compared to the equivalent non-rotating polytropic models (for details
see Sect.~\ref{sect:SpinBH} and the end of  Sects.~\ref{sect:orbit}
and~\ref{sect:dmass}). 
It is beyond the scope of this paper to map out the differences in
detail over the full parameter space, but this will be the topic of a
subsequent investigation.  
We note that Etienne et al.~(2009) report a fairly high value for the
torus mass ($0.2\,M_\odot$) in their $Q\!=\!3$ model with a spinning BH 
($a\!=\!0.75$).

Our values for the final spin parameter of the BH (0.39, 0.43, 0.56) 
match the values obtained by Shibata et al.~(2009; their Table~III)
fairly well, although their values are systematically larger (0.42,
0.56, 0.68, respectively). 

In our pseudo-Newtonian simulations, the mass that assembles
into a torus or is ejected, shows a very strong sensitivity
to the inclusion of the angular momentum gain by the BH because of
the accretion of matter in the merger. 
This effect, however, should be generically included in the
relativistic models of Shibata et al.~(2009) and Etienne et al.~(2009).

Since the existence of an ISCO is accounted for by the BH potentials
used in our simulations, the tendency towards reaching higher torus
and ejecta masses in our models than relativistic simulations (with
similar system parameters) requires a different explanation. 
We hypothesise that the difference could be a consequence of the
stronger self-gravity of relativistically described neutron stars 
compared to the Newtonian objects considered in our simulations. 
We note how so much more closely the tidal disruption limits of the
pseudo-TOV models of Ferrari et al.~(2009) shown in their Fig.~1,
match the full GR results than the purely Newtonian models.
The deeper relativistic potential might impede the disruption of the
neutron star and the formation of a tidal arc, so that a greater
fraction of the star is directly swallowed by the BH.


\section{Conclusions\label{sect:conclude}}

We reach the following conclusions.

\begin{enumerate}
\item We do not note any fundamental and discontinuous difference in the 
  dynamics of the models in the ISCO regime as opposed to the `mass
  shedding' (= tidal disruption) regime. The variation in masses that
  are not instantly accreted by the BH is
  continuous (Figs.~\ref{fig:ISCO.MS} and~\ref{fig:ISCO.MS.Shi}).
\item In all cases the NS is initially tidally stretched into an arc,
  albeit of very different mass in each of the various cases. 
\item For models without BH spin, the onset of orbital widening is
  noted during the phase of tidal shredding of the NS but is only
  short-lived.  
\item In one case of a model including the BH spin, we see the
  formation of a `mini' NS on an extended elliptic orbit, followed by
  episodic mass transfer, and a final tidal disruption. 
\item In mergers with compact neutron stars (${\cal C}\!=\!0.180$),
  a very small amount, less than $10^{-2}M_\odot$, remains in the
  surroundings of the BH.
\item Only in mergers of low mass BHs with mass ratio $Q\!=\!2$ or
  $Q\!=\!3$, does a closed accretion torus -- as opposed to an
  open arc -- form.
\item Mergers with neutron stars of compactness ${\cal C}\!=\!0.145$
  or ${\cal C}\!=\!0.160$ produce a significant amount of material
  (order of $0.1M_\odot$) in the surroundings of the BH.
 \item Effects due to BH spin-up are important and significantly change
   the results. We note an increase in mass of the material
   surrounding the BH of up to 0.2--0.3~$M_\odot$.
\end{enumerate}

\begin{acknowledgements}
We would like to thank the anonymous referee for carefully reading the
manuscript and providing helpful and detailed suggestions for 
improvement of this work. 
MR would like to thank the Max-Planck-Institut f\"ur Astro\-physik
for the kind hospitality during a sabbatical visit, where this
work was accomplished.
We thank A.~Bauswein for stimulating and helpful discussions.
The calculations were performed at the Rechenzentrum Garching (RZG).
The project was also supported by the Deutsche Forschungsgemeinschaft
through the Transregional Collaborative Research Centers SFB/TR~27 
``Neutrinos and Beyond'' and SFB/TR~7 ``Gravitational Wave Astronomy'',
and the Clus\-ter of Excellence EXC~153 
({\tt http://www.universe-cluster.de}) ``Origin and Structure of the 
Universe''.

\end{acknowledgements}


\begin{thebibliography}{}

\bibitem[2009]{abr09} Abramowicz, M.A. 2009, A\&A, 500, 213

\bibitem[1996]{art96} Artemova, I.V., Bj\"ornsson, G. \&
   Novikov, I.D. 1996, ApJ, 461, 565

\bibitem[1984]{col84} Colella, P. \& Woodward, P.R. 1984,
   J.~Comput.Phys., 54, 174

\bibitem[2005]{dav05} Davies, M.B., Levan, A.J. \& King, A.R.
   2005, MNRAS, 356, 54

\bibitem[2008]{due08} Duez, M., Foucart, F., Kidder, L., 
   Pfeiffer, H., Scheel, M. \& Teukolsky, S. 2008, 
   Phys.~Rev.~D, 78, 104015 

\bibitem[2009]{due09} Duez, M. 2009, arXiv:0912.3529

\bibitem[2008]{eti08} Etienne, Z.B., Faber, J.A., Liu, Y.T., 
   Shapiro, S.L., Taniguchi, K. \& Baumgarte, T.W. 2008,
   Phys.~Rev.~D, 77, 084002

\bibitem[2009]{eti09} Etienne, Z.B., Liu, Y.T., Shapiro, S.L.
   \& Baumgarte, T.W. 2009, Phys.~Rev.~D, 79, 044024

\bibitem[2009]{fab09} Faber, J. 2009, Class.~Quantum Grav., 26, 
   114004 

\bibitem[2009]{fer09} Ferrari, V., Gualtieri, L., \& Pannarale, F.
   2009, Class.~Quant.~Grav., 26, 125004

\bibitem[1999]{jan99} Janka, H.-Th., Eberl, T., Ruffert, M.,
   \& Fryer C.L. 1999, ApJ, 527, L39

\bibitem[1999]{lee99} Lee, W.H. \& Klu\'zniak, W. 1999, 
   ApJ, 526, 178

\bibitem[2005]{lee05} Lee, W.H., Ramirez-Ruiz, E. \& Granot, J.
   2005, ApJL, 630, L165

\bibitem[2005]{mil05} Miller, M.C. 2005, ApJ, 626, L41

\bibitem[2006]{oec05} Oechslin, R. \& Janka, H.-Th. 2006, 368, 1489

\bibitem[1980]{pac80} Paczy\'nski, B. \& Wiita, P. 1980,
   A\&A, 88, 32

\bibitem[1998]{por98} Portegies Zwart, S.F. 1998, ApJ, 503, L53

\bibitem[2008]{ran08} Rantsiou, E., Kobayashi, S., Laguna, P.
   \& Rasio, F.A. 2008, A\&A, 680, 1326

\bibitem[1999]{ras99} Rasio, F.A. \& Shapiro, S.L. 1999, 
   Class.~Quant.~Grav., 16, R1

\bibitem[2004]{ros04} Rosswog, S., Speith, R. \& Wynn, G.A. 
   2004, MNRAS, 351, 1121

\bibitem[1992]{ruf92} Ruffert, M. 1992, A\&A, 265, 82

\bibitem[2001]{ruf01} Ruffert, M. \& Janka, H.-Th. 2001, 
   A\&A, 380, 544

\bibitem[1996]{ruf96} Ruffert, M., Janka, H.-Th. \&
   Schaefer, G. 1996, A\&A, 311, 532

\bibitem[1997]{ruf97} Ruffert, M., Janka, H.-Th., Takahashi, K. \&
   Schaefer, G. 1997, A\&A, 319, 122

\bibitem[2006]{set06} Setiawan, S., Ruffert, M. \& Janka, H.-Th.
   2006, A\&A, 458, 553

\bibitem[1998]{she98} Shen, H., Toki, H., Oyamatsu, K. \& 
   Sumiyoshi, K., 1998, Nucl.~Phys.~A, 637, 435 

\bibitem[2009]{shi09} Shibata, M., Kyutoku, K., Yamamoto, T.
   \& Taniguchi, K. 2009, Phys.~Rev.~D, 79, 044030

\bibitem[2008]{tan08} Taniguchi, K., Baumgarte, T.W., 
   Faber, J.A. \& Shapiro, S.L. 2008, Phys.~Rev.~D, 77, 044033

\end{thebibliography}
\end{document}